\documentclass[onecolumn]{emulateapj}
\usepackage{amssymb,amsmath,xcolor}

\newcommand{\pam}{\textsf{PAMELA}}
\newcommand{\deu}{\textsuperscript{2}H}
\newcommand{\prot}{\textsuperscript{1}H}
\newcommand{\het}{\textsuperscript{3}He}
\newcommand{\hef}{\textsuperscript{4}He}

\submitted{}

\begin{document}

\title{Measurements of Cosmic-Ray Hydrogen and Helium Isotopes with the
\pam\ experiment}

  \author{
O. Adriani$^{1,2}$, G. C. Barbarino$^{3,4}$,
  G. A. Bazilevskaya$^{5}$, R. Bellotti$^{6,7}$, 
M. Boezio$^{8}$, \\
E. A. Bogomolov$^{9}$, M. Bongi$^{1,2}$, V. Bonvicini$^{8}$,
S. Bottai$^{2}$, A. Bruno$^{6,7}$, F. Cafagna$^{7}$, \\
D. Campana$^{4}$,
P. Carlson$^{13}$, 
M. Casolino$^{11}$, G. Castellini$^{14}$, C.~De~Donato$^{10}$, \\
C. De Santis$^{11}$, N. De
Simone$^{10}$, V. Di Felice$^{10}$, V. Formato$^{8,\dagger}$, 
 A. M. Galper$^{12}$, \\
A. V. Karelin$^{12}$, 
S. V. Koldashov$^{12}$, S. Koldobskiy$^{12}$, S. Y. Krutkov$^{9}$, \\
A. N. Kvashnin$^{5}$, A. Leonov$^{12}$, 
V. Malakhov$^{12}$, L. Marcelli$^{11}$,  M.~Martucci$^{11,15}$, \\
A. G. Mayorov$^{12}$, W. Menn$^{16}$, 
M Merg\`{e}$^{10,11}$,
V. V. Mikhailov$^{12}$, 
E. Mocchiutti$^{8}$, \\ 
A. Monaco$^{6,7}$,  
N. Mori$^{2}$, R.~Munini$^{8,17}$,
G. Osteria$^{4}$,
F. Palma$^{10,11}$, 
B.~Panico$^{4}$,  \\
P. Papini$^{2}$, M. Pearce$^{13}$, P. Picozza$^{10,11}$, 
M. Ricci$^{15}$,
S. B. Ricciarini$^{14}$, \\
R. Sarkar$^{18,*}$, V.~Scotti$^{3,4}$, M. Simon$^{16}$,
 R. Sparvoli$^{10,11}$, P. Spillantini$^{1,2}$, \\
Y. I. Stozhkov$^{5}$, A. Vacchi$^{8,19}$, 
E. Vannuccini$^{2}$, G. Vasilyev$^{9}$, S. A. Voronov$^{12}$, \\
 Y. T. Yurkin$^{12}$, 
 G. Zampa$^{8}$, N. Zampa$^{8}$}
\affil{$^{1}$University of Florence, Department of Physics, I-50019 Sesto Fiorentino, Florence, Italy}
\affil{$^{2}$INFN, Sezione di Florence, I-50019 Sesto Fiorentino, Florence, Italy}
\affil{$^{3}$University of Naples ``Federico II'', Department of Physics, I-80126 Naples, Italy}
\affil{$^{4}$INFN, Sezione di Naples,  I-80126 Naples, Italy}
\affil{$^{5}$Lebedev Physical Institute, RU-119991, Moscow, Russia}
\affil{$^{6}$University of Bari, Department of Physics, I-70126 Bari, Italy}
\affil{$^{7}$INFN, Sezione di Bari, I-70126 Bari, Italy}
\affil{$^{8}$INFN, Sezione di Trieste, I-34149 Trieste, Italy}
\affil{$^{9}$Ioffe Physical Technical Institute,  RU-194021 St. Petersburg, Russia}
\affil{$^{10}$INFN, Sezione di Rome ``Tor Vergata'', I-00133 Rome, Italy}
\affil{$^{11}$University of Rome ``Tor Vergata'', Department of Physics,  I-00133 Rome, Italy}
\affil{$^{12}$National Research Nuclear University MEPhI, RU-115409 Moscow}
\affil{$^{13}$KTH, Department of Physics, and the Oskar Klein Centre for Cosmoparticle Physics, AlbaNova University Centre, SE-10691 Stockholm, Sweden}
\affil{$^{14}$IFAC, I-50019 Sesto Fiorentino, Florence, Italy}
\affil{$^{15}$INFN, Laboratori Nazionali di Frascati, Via Enrico Fermi 40, I-00044 Frascati, Italy}
\affil{$^{16}$Universit\"{a}t Siegen, Department of Physics, D-57068 Siegen, Germany}
\affil{$^{17}$University of Trieste, Department of Physics, I-34147 Trieste, Italy}
\affil{$^{18}$Indian Centre for Space Physics, 43 Chalantika, Garia Station Road, Kolkata 700084, West Bengal, India}
\affil{$^{19}$University of Udine, Department of Mathematics and Informatics, I-33100 Udine, Italy}
\affil{$^{\dagger}$Now at INFN, Sezione di Perugia, I-06123 Perugia, Italy}
\affil{$^{*}$Previously at INFN, Sezione di Trieste, I-34149 Trieste, Italy}

\begin{abstract}
The cosmic-ray hydrogen and helium (\prot,\deu,\het,\hef) isotopic composition has been measured with the  satellite-borne experiment \pam, which was launched into low-Earth orbit on-board the Resurs-DK1 satellite on June 15\textsuperscript{th} 2006.
The rare isotopes \deu\ and \het\ in cosmic rays are believed to originate mainly from the interaction of high energy protons and helium with the galactic interstellar medium.
The isotopic composition was measured between 100 and 1100 MeV/n for hydrogen and between 100 and 1400 MeV/n for helium isotopes using two different detector systems over the 23\textsuperscript{rd} solar minimum from July 2006 to December 2007.
\end{abstract}


\section{Introduction}

The rare isotopes \deu\ and \het\ in cosmic rays
are generally believed to be of secondary origin, 
resulting mainly from the nuclear interactions of primary cosmic-ray protons and \hef\
with the interstellar medium.
The spectral shape and composition of the secondary isotopes is therefore 
completely determined by the source spectrum of the parent elements
and by the propagation process. Measurements of the secondary isotopes
spectra are then a powerful tool to constrain the parameters of the
galactic propagation models \citep{2007ARNPS..57..285S,2012Ap&SS.342..131T,2012A&A...539A..88C,1995ApJ...441.209}.

The first measurements of hydrogen and helium isotopes became available in the seventies \citep{1975ICRC....1..319G,1975ApJ...202..265G,1976ApJ...206..616M,1978ApJ...221.1110L},  
but limited to energies below 100 MeV/n. The identification, especially at
energies greater than 100 MeV/n is quite difficult due to the high
experimental mass resolution required to distinguish the secondary
nuclei from the abundant background of primaries. In the eighties and nineties there have been several measurements of stratospheric balloon experiments using superconducting magnetic spectrometers
\citep{2002ApJ...564..244W,1998ApJ...496..490R_red,1995ICRC....2..630W,1991ApJ...380..230W,1993ApJ...413..268B_red} 
and also a measurement from AMS-01 \citep{2011ApJ...736..105A_red} in space.
While the mass resolution of the balloon experiments was usually quite good, the residual atmosphere above the instruments caused a non-negligible background of secondary particles. The atmospheric background estimation is subject to large uncertainties (e.g. the limited knowledge of isotope production cross sections).

The results presented in this paper are based on data gathered between July 2006 and December 2007 with the \pam\ satellite experiment. \pam\ has been put in a polar elliptical orbit at an altitude between $\sim 350$ and $\sim 600$ km with an inclination of $70^\circ$ as part of the Russian Resurs-DK1 spacecraft. Due to the low-earth orbit the measurements are performed in an environment free from the background induced by interactions of cosmic rays within the atmosphere.
The month of December 2006 was discarded from the dataset to avoid possible biases from the solar particle events that took place during the 13\textsuperscript{th} and 14\textsuperscript{th} of December. During a total acquisition time of 528 days about $10^9$ triggered events were recorded, $5.8 \cdot 10^6$ hydrogen nuclei were selected in the energy interval between 100 and 1100 MeV/n and $1.6 \cdot 10^6$ helium nuclei between 100 and 1400 MeV/n.

This is the second paper on isotopes from the \pam\ instrument. The first paper  \citep{2013ApJ...770..2} dealt with the same isotopes \prot, \deu, \het, \hef\ but used solely the combination of velocity measurement provided by the time-of-flight system in combination with the momentum measurement of the magnetic spectrometer. 
In this work here we employed a more advanced and a more  comprehensive analysis. In detail: a more complete and elaborated fitting procedure was employed, combined with a more stringent selection on cuts and on efficiencies.  In addition we made use of the multiple energy loss measurements provided by the 44 planes of the imaging calorimeter.  This not only allowed a cross check between the two techniques of isotopic separation within the \pam\  instrument (ToF and multiple $dE/dx$ versus rigidity), the multiple $dE/dx$ technique also allowed to extend the measurements for isotopes to higher energies: For hydrogen isotopes the highest energy bin is now at 1035 MeV/n (instead of 535 MeV/n), while for helium it is now at 1297 MeV/n (instead of 823 MeV/n). Additionally the previous results \citep{2013ApJ...770..2} were revised in this improved analysis.

\section{The \pam\ apparatus}

A schematic view of the \pam\ detector system \citep{2007APh....27..296P} is shown in Figure 1. The design was chosen to meet the main scientific goal of precisely measuring the light
components of the cosmic ray spectrum in the energy range starting from tens of MeV up to 1 TeV (depending on particle species), with a particular focus on antimatter. Thus the design is optimized for $Z=1$ particles and a high lepton/hadron discrimination power.

\begin{figure}[t]
    \centering
    \epsscale{0.5}
    \plotone{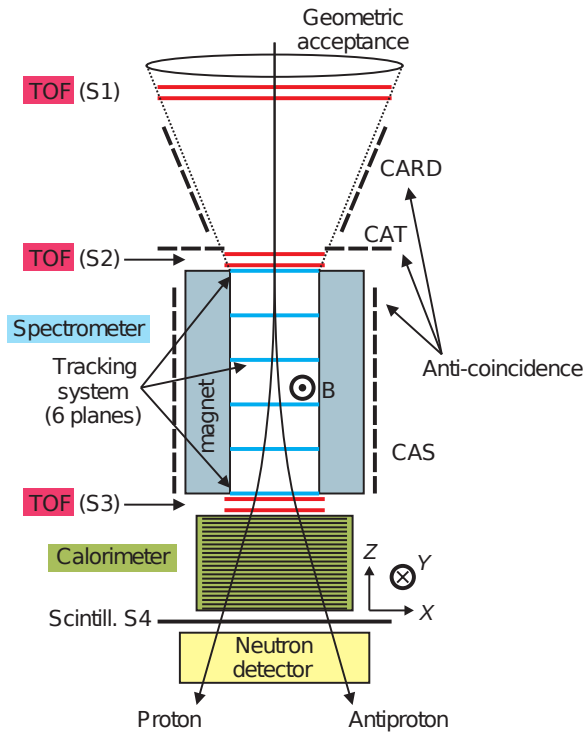}
    \caption{Scheme of the detectors composing the \pam\ satellite experiment.}
    \label{im:pamela}
\end{figure}

The instrument core is a permanent magnet with a silicon microstrip tracker. 
The design of the permanent magnet provides an almost uniform magnetic field of 0.45 T inside the magnetic cavity. Six layers of 300 $\mu$m thick double-sided microstrip silicon detectors are used to measure particle deflection with $\sim3$ $\mu$m  and $\sim11$ $\mu$m precision (measured with beam tests and flight data) in the bending and non-bending views, respectively.
Due to the small size the amount of material inside the magnetic cavity can be kept to a minimum, only the six layers of silicon without any need for support structure. 
The MDR (Maximum Detectable Rigidity) of the magnetic spectrometer is about 1 TV.

The Time-of-Flight (ToF) system comprises six layers of fast plastic
scintillators arranged in three planes (S1, S2 and S3). Each detector layer is segmented into strips, placed in alternate layers orthogonal to each other. 
Using different combinations of layers, the ToF system can provide 12 measurements of the particle velocity, $\beta = v/c$, by using a weighted mean technique an overall value for  $\beta$ is calculated from these measurements  . 
The overall time resolution of the ToF system is about 250 ps for $Z = 1$ particles and about 100 ps for $Z = 2$ particles. This allows albedo particles crossing \pam\ from bottom to top to be discarded by requiring a positive $\beta$. The TOF scintillators can also identify the absolute particle charge up to oxygen by means of the six independent ionization measurements.

The silicon-tungsten electromagnetic sampling calorimeter comprises 44 single-sided silicon planes interleaved with 22 plates of tungsten absorbers. The calorimeter is mounted below the spectrometer, its primary use is for lepton/hadron separation \citep{2002NIMPA.487..407B}. Each tungsten layer has a thickness of 0.74 radiation lengths (2.6 mm) and it is sandwiched between two printed circuit boards, which house the silicon detectors as well as the frontend and digitizing electronics. Each silicon plane consists of 3x3, 380 $\mu$m thick, 8x8 cm$^2$ detectors, segmented into 32 strips with a pitch of 2.4 mm. The orientation of the strips for two consecutive silicon planes is shifted by 90 degrees, thus providing 2-dimensional spatial information. The total depth of the calorimeter is 16.3 radiation lengths and 0.6 nuclear interaction lengths. 
Below the calorimeter there is a shower tail catcher scintillator (S4) and a neutron detector which help to increase hadron/lepton separation. 
The tracking system and the upper ToF system are shielded by an anticoincidence system (AC) made of plastic scintillators and arranged in three sections (CARD, CAT, and CAS in Fig. \ref{im:pamela}), which allows to detect during offline data analysis the presence of secondary particles generating a false trigger or the signature of a primary particle suffering an inelastic interaction.
The total weight of \pam\ is 470 kg and the power consumption is 355 W. A more detailed description of the instrument can be found in \cite{2007APh....27..296P}.

\section{Data analysis}
\subsection{Event selection}   \label{sec:event_selection}

\begin{figure}[t]
    \centering
    \plotone{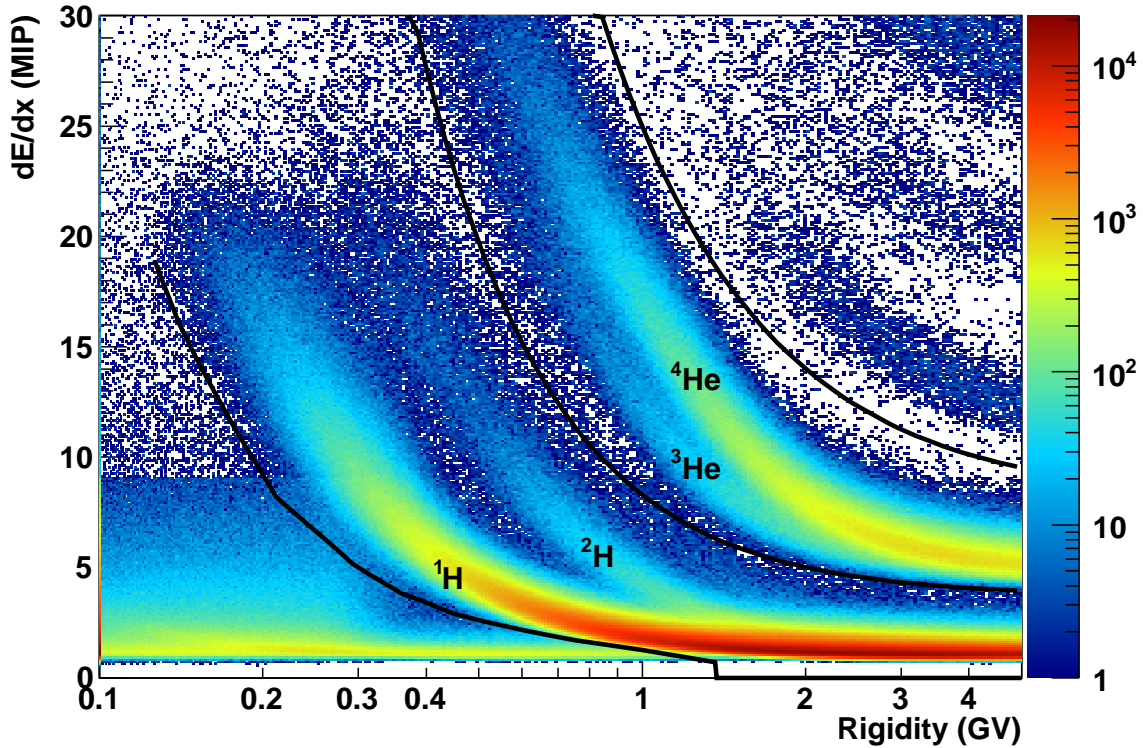}
    \caption{
		Ionization loss ($dE/dx$ in MIP, energy loss of minimum ionizing particles) in the silicon detectors of the tracking system as a function of  reconstructed rigidity. The black lines represent the selection for $Z=1$ or $Z=2$ nuclei.}
    \label{im:dedx}
\end{figure}

Each triggered event had to fulfill several criteria to be used for further analysis. The requirements are identical to the selection in \cite{2013ApJ...770..2} and we refer to that paper for more details:

\begin{itemize}

\item Event quality selections: We have selected events that do not produce secondary
particles by requiring a single track fitted within the spectrometer
fiducial acceptance and a maximum of one paddle hit in the two top scintillators of the ToF system. The analysis procedure was similar to previous work on high energy proton and helium fluxes \citep{2011Sci...332...69A}.

\item Galactic particle selection:
Galactic events were selected by imposing that the lower edge of the rigidity bin to which the event belongs exceeds the critical rigidity, $\rho_c$, defined as 1.3 times the cutoff rigidity $\rho_{SVC}$ computed in the
St\"{o}rmer vertical approximation~\citep{1987PEPI...48..200S}
as $\rho_{SVC}=14.9/L^2$, where L is the McIlwain L-shell parameter obtained by using the Resurs-DK1 orbital information and the IGRF magnetic field model \citep{2005EP&S...57.1135M}.

\item Charge selection:
The charge identification uses the ionization measurements provided by the silicon sensors of the magnetic spectrometer. Depending on the number of hit sensors there can be up to 12 $dE/dx$ measurements, the arithmetic mean of those measurements is shown in Fig. \ref{im:dedx} as a function of the rigidity. The actual selection of $Z=1$ or $Z=2$ particles is depicted by the solid lines. A similar figure has been already shown in \cite{2013ApJ...770..2}
but was kept in this paper to help the reader.

\end{itemize}

\subsection{Isotope separation in the \pam\ instrument}

In each sample of $Z = 1$ and $Z=2$ particles an isotopic separation at fixed rigidity is possible since the mass of each particle follows the relation

\begin{equation}
  m = \frac{RZe}{\gamma\beta c}
\label{eq:mass_calculation}
\end{equation}

where $R$ is the magnetic rigidity, $Z e$ is the particle charge, and $\gamma$ stands for the Lorentz factor. The particle velocity $\beta$ can either be provided directly from the timing measurement of the ToF system, or indirectly from the energy loss in the calorimeter, which follows $\beta$ via the Bethe-Bloch formula $dE/dx \propto \frac{Z^2}{\beta^2}$ (neglecting logarithmic terms).

\subsubsection{Isotope separation using ToF vs. rigidity}

\begin{figure}[t]
    \centering
    \plotone{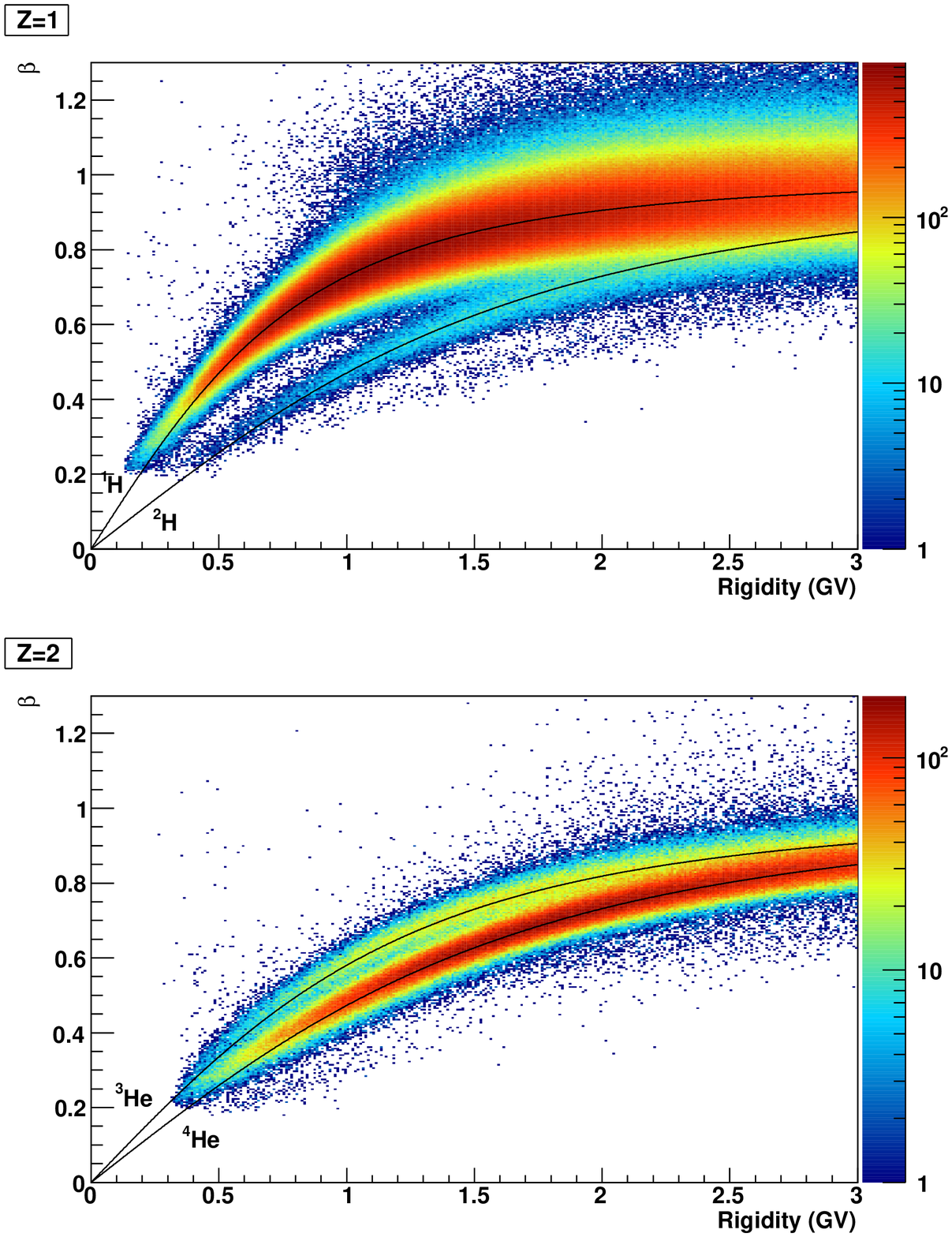}
    \caption{ $\beta$ vs. rigidity for $Z=1$ (top) and $Z=2$ (bottom) particles. The black lines were calculated for each isotope using Eq. ~\ref{eq:mass_calculation}.}
    \label{im:beta_r}
\end{figure}

For the ToF analysis we can use directly the $\beta$ provided by the timing measurement. In Fig.~\ref{im:beta_r} we show $\beta$ vs. the particle rigidity for $Z = 1$ and $Z = 2$ data. The black lines in the figure represent the expectations for each isotope. A similar figure has been already shown in \cite{2013ApJ...770..2} but was kept in this paper for an easier comparison with the alternative identification method using the calorimeter.

Since the particle mass is calculated using Eq. ~\ref{eq:mass_calculation}, misidentified \hef\ wrongly reconstructed as $Z=1$ particles could result in a significant contamination to the \deu\ sample. However, the amount of misidentified helium was found to be negligible, see \cite{2013ApJ...770..2} for more details.

\subsubsection{Isotope separation using multiple $dE/dx$ in the calorimeter vs. rigidity}\label{sec:calo_principle}

The isotopic analysis of nuclei with the calorimeter is restricted to events which do not interact inside the calorimeter.
To check if an interaction occurs, we derived in each silicon layer a) the total energy detected ($q_{tot}$) and b) the energy deposited in the strip closest to the track and the neighbouring strip on each side ($q_{track}$). In the ideal case the fraction of $q_{track}/q_{tot}$ will be equal to one, a value less then one means that strips outside the track were hit.
Starting from the top of the calorimeter we calculated $\Sigma q_{track}/ \Sigma q_{tot}$ at each layer, as long as this value is greater than 0.9, we used these layers for further analysis. A value of 0.9 was chosen since it was found to give a good compromise between high efficiency and rejection of interactions.
In this way we can make use of slow particles, which stop early in the calorimeter, particles which interact somewhere, but also all clean events with the particle fully traversing the calorimeter.
In the single silicon layer the energy loss distributions shows a Landau tail which degrades the resolution of the dE/dx measurement. Using a truncation method, the 50\% of samples with larger pulse amplitudes were excluded before taking the mean of the dE/dx measurements, thus reducing the effect of the Landau tail.
We put an energy dependent lower limit on the number of layers after the 50\% truncation, requiring at least 5 measurements at 1 GV, going up to 10 layers at 3 GV. With this requirement the lower energy limit of our analysis is around 200 - 300  MeV/n (the energy to fully penetrate the calorimeter is much higher, about 400 - 500 MeV/n).

\begin{figure}[t]
    \centering
    \plotone{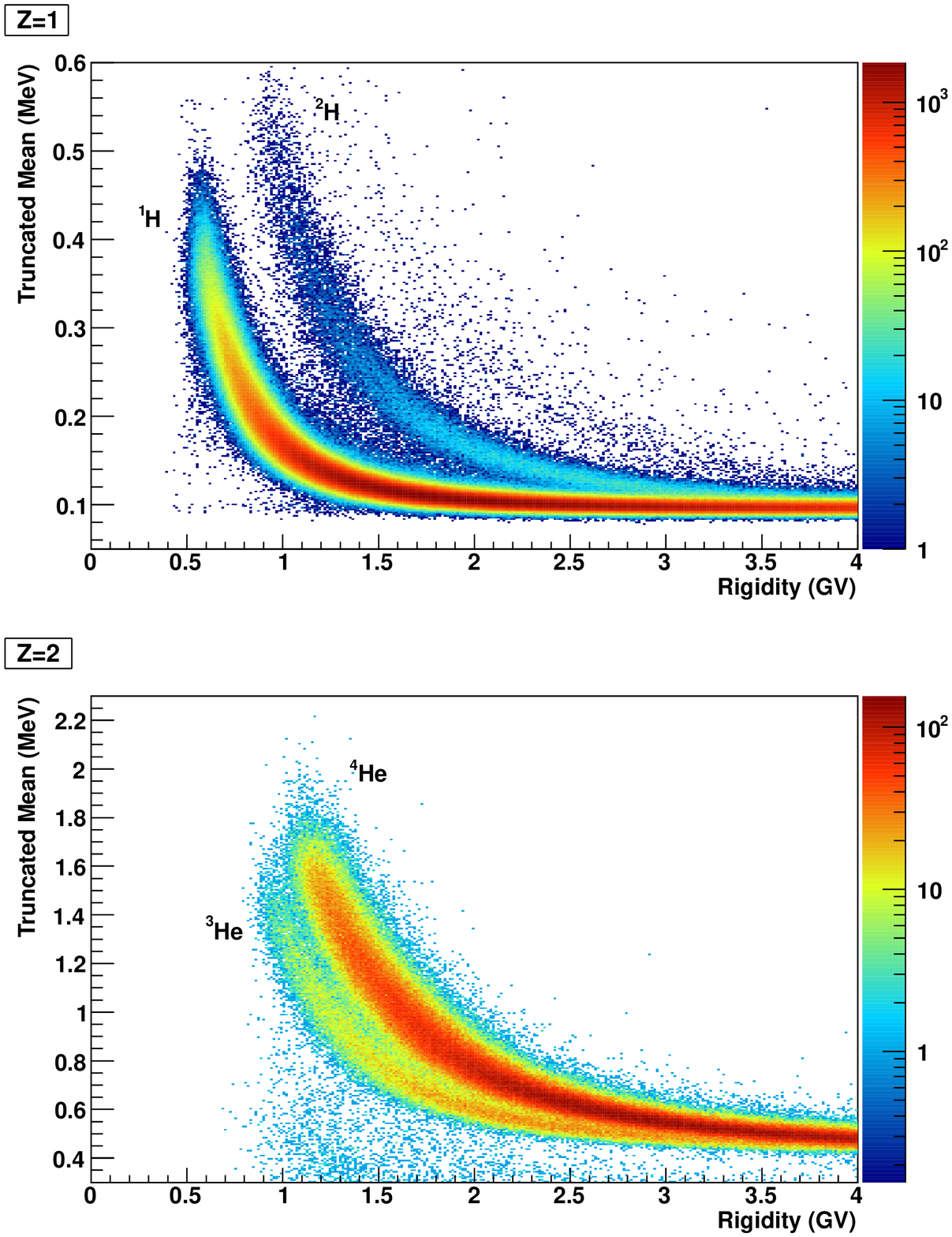}
    \caption{Mass separation for $Z=1$ (top) and $Z=2$ (bottom) particles using the ``truncated mean''-method. }
    \label{im:calo_tm_r}
\end{figure}

In Fig.~\ref{im:calo_tm_r} we show the mean dE/dx for each event vs. the rigidity measured with the magnetic spectrometer for $Z = 1$ and $Z = 2$ particles. The energy loss in MeV was calculated from the measurement in MIP using a conversion factor (the most probable energy loss of a minimum ionizing particle traversing 380 $\mu$m of silicon, derived by simulation). In both plots the isotopic separation is clearly visible.
It is worthwhile to mention that the selection procedure described above 
can be done in different ways but we found that we achieved the best results particularly on the corresponding efficiency  (see section \ref{sec:flux_determination}) by the method described above. 

The use of the calorimeter for isotope separation provides us with another advantage. Contrary to the $\beta$ vs rigidity technique with the ToF, by using the multiple $dE/dx$ vs. rigidity technique a misidentified \hef\ wrongly reconstructed as $Z=1$ particles would not result in a contamination to the \deu\ sample, since the $dE/dx$ is different. This applies also for the $Z=1$ contamination and contamination by heavier nuclei in the $Z=2$ sample. This allows a valuable crosscheck between the results from ToF and calorimeter in the energy regime where the two measurements overlap.

\subsubsection{Measured Mass resolution in the \pam\ instrument and comparison with expectations} \label{sec:mass_resolution}

Theoretically, there are three independent contributions to the mass resolution in a magnet spectrometer similar to \pam: the bending power of the magnetic spectrometer coupled with the intrinsic limits of spatial resolution which the tracking detectors provide, the precision of the velocity measurement (given either by timing or by measuring the energy loss), and the multiple scattering of the particle along its path in the bending area of the magnet. These three independent contributions can be expressed by the following equation:
\begin{equation}
  dm =  m \sqrt { \gamma ^4 \left(\frac{d\beta}{\beta}\right)^2  + \left(\frac{R}{MDR_{spec}}\right)^2 + \left(\frac{R}{MDR_{cou}}\right)^2  }
 \label{eq:mass_resolution}
\end{equation}

where $\gamma$ is the Lorentz factor, $d\beta/\beta$ is the relative error in the
velocity measurement, $R/MDR_{spec}$ stands for the contribution solely given by the magnetic spectrometer, and the last term stands for multiple coulomb scattering.
The last two terms of  Eq. ~\ref{eq:mass_resolution} can be expressed by an overall $MDR_{tot}$:

\begin{equation}
 \left(\frac{1}{MDR_{tot}}\right)^2  = \left(\frac{1}{MDR_{spec}}\right)^2 + \left(\frac{1}{MDR_{cou}}\right)^2  
 \label{eq:eff_mdr}
\end{equation}

For \pam\ the overall momentum resolution of the magnetic spectrometer (thus $MDR_{tot}$) has been measured in beam tests at CERN \citep{2007APh....27..296P}. From these measurements at high energies, where the contributions from multiple scattering is negligible, one can derive that the $MDR_{spec}$ of the \pam\ spectrometer has a value of about 1 TV for $Z = 1$ particles. The contribution of $MDR_{spec}$ to the overall mass resolution is therefore negligible for the particles we are analyzing (up to some GV). At low rigidities the contribution from multiple scattering is however the dominant effect. Its value is inverse proportional to the bending power of the magnet ($\int B \cdot dl$) and direct proportional to the amount of matter traversed along the bending part of the track.
For an experiment using a permanent magnet in combination with silicon strip detectors \pam\ shows a very good momentum resolution, due to its strong magnetic field of 0.45 T combined with a low amount of material (only the six silicon detectors, each 300 $\mu$m, giving a grammage of 0.42 g/cm$^2$) in the magnetic cavity. The overall momentum resolution was measured at CERN to have a minimum of about 3.5 \% at 8 GV, increasing to 5\% at 1 GV \citep{2007APh....27..296P}. 
Using Eq.~\ref{eq:eff_mdr} one can derive the respective values for $MDR_{cou}$.

\begin{figure}[t]
    \centering
    \plottwo{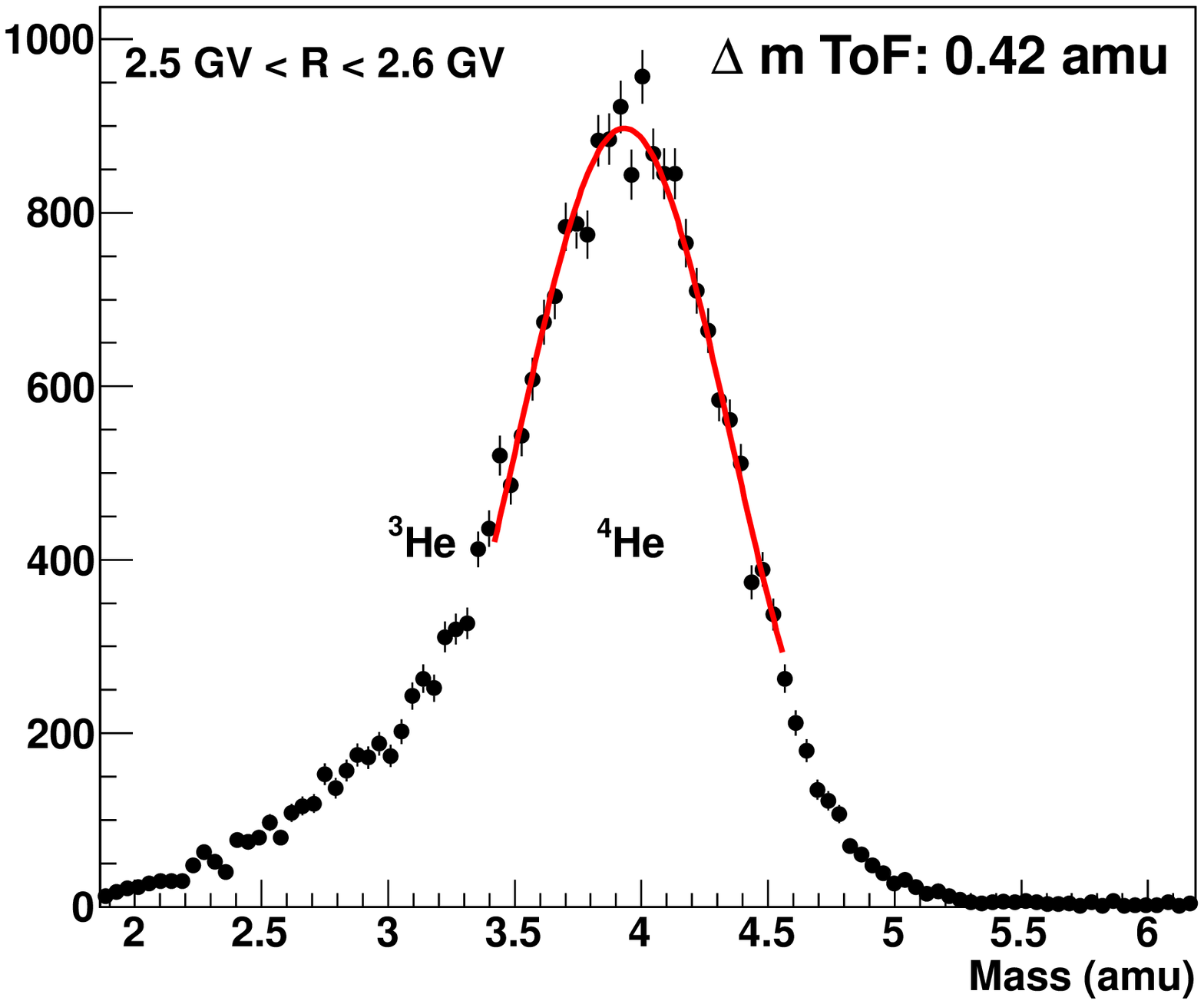}{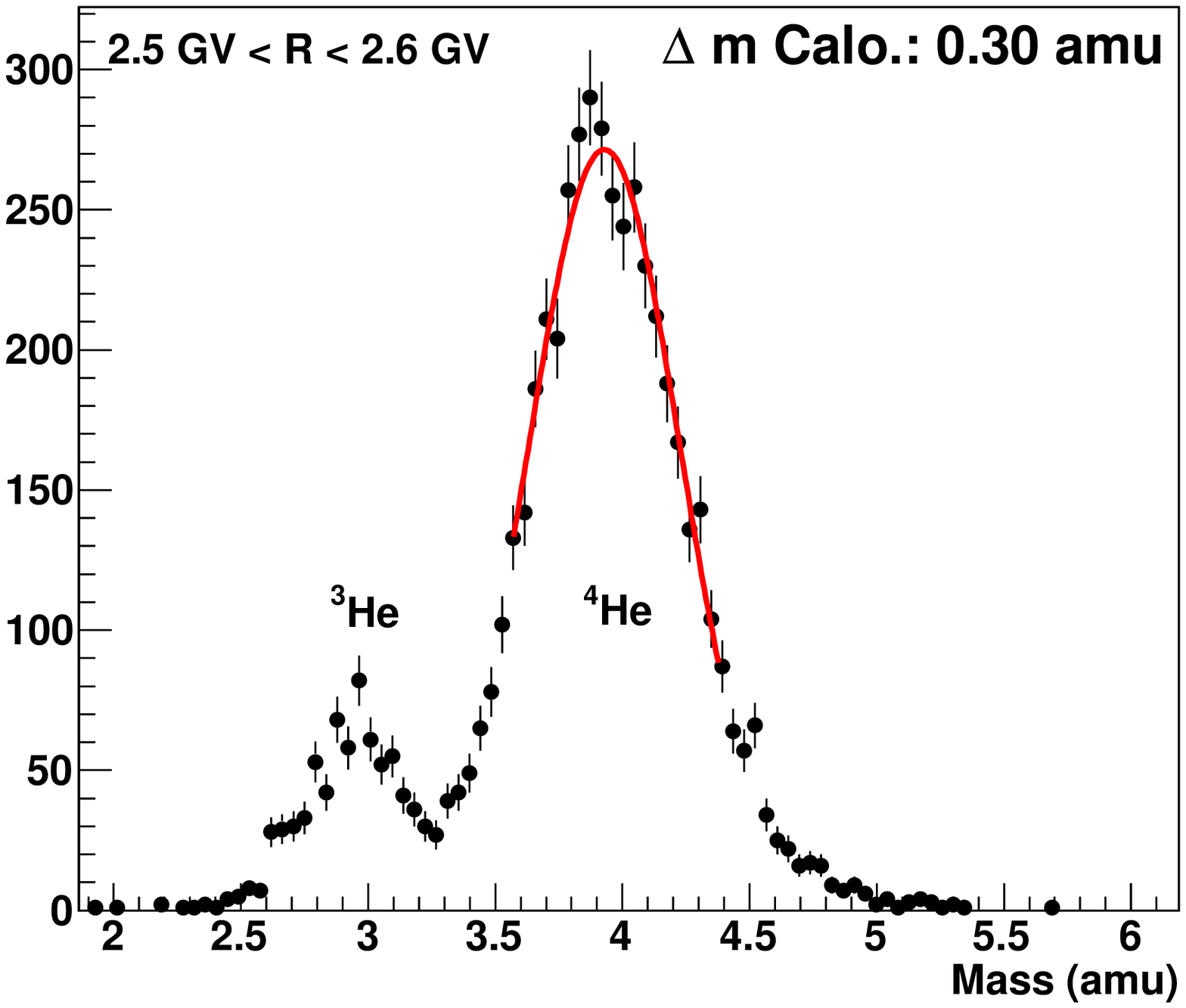}
       \caption{Examples mass distributions for helium in the 2.5 - 2.6 GV rigidity range for ToF (left) and Calorimeter (right)}.
    \label{im:he_mass_histograms}
\end{figure}

By using the $\beta$-rigidity technique (see Fig.~\ref{im:beta_r}) Eq.~\ref{eq:mass_calculation} directly provides the mass of the particle and a corresponding mass (amu) histogram for helium in the rigidity range from 2.5 to 2.6 GV is shown in Fig.~\ref{im:he_mass_histograms} (left).
When using the multiple $dE/dx$ versus rigidity technique (see Fig.~\ref{im:calo_tm_r}) either a mass of three or four amu was allocated to the corresponding peaks in the histogram. By scaling between these positions linearly one obtained the mass distribution which we show also for helium and for the same rigidity range of 2.5 to 2.6 GV in Fig.~\ref{im:he_mass_histograms} (right).
We fitted a gaussian to the \hef\ peak and used the standard deviation of the gaussian as the mass resolution, in this example 0.42 amu with the ToF, while it is 0.30 amu for the calorimeter. Thus the use of the multiple $dE/dx$ in the calorimeter stack provided a better mass resolution than the direct measurement of $\beta$ with the ToF for this rigidity interval.

\begin{figure}[t]
\centering
\plotone{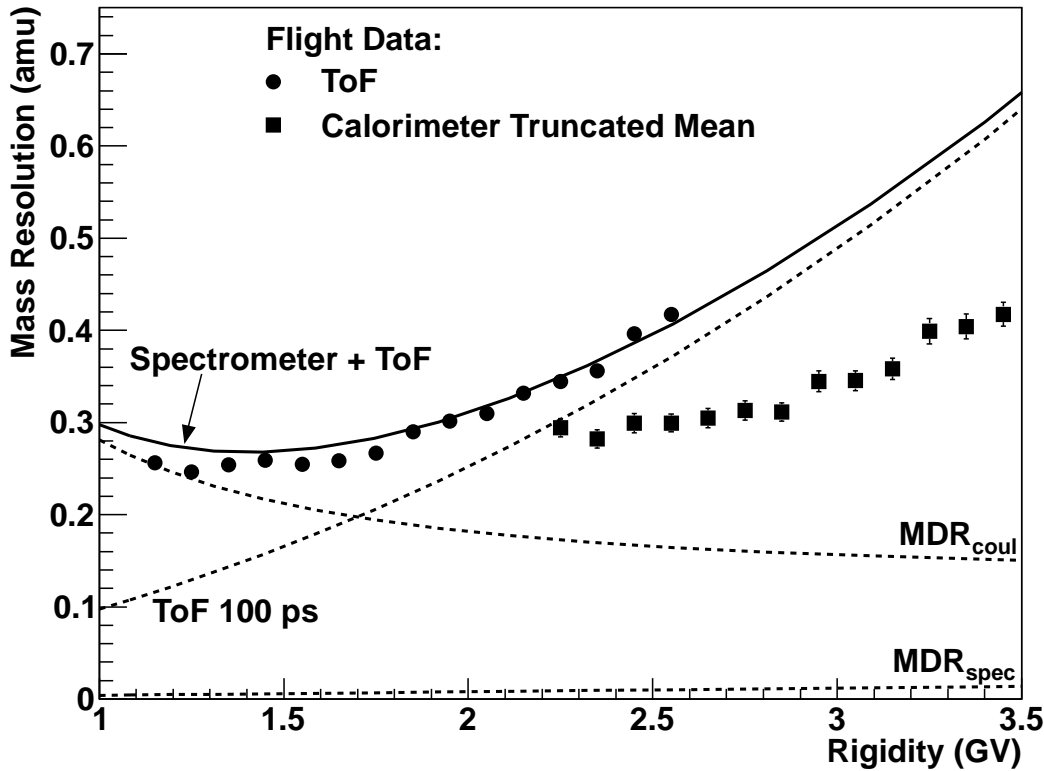}
\caption{\label{im:he_mass_resolution}Measured \hef\ mass resolution for the ToF (circles) and the calorimeter using the ``truncated mean'' method (squares). If the error bars are not visible, they lie inside the data points. The dashed lines show the calculated independent contributions (rigidity ($MDR_{spec}$), multiple scatter ($MDR_{cou}$), and velocity via ToF (time resolution 100 ps)), while the solid lines shows the overall mass resolution for this combination.}
\end{figure}

We repeated this procedure for a number of rigidity intervals and derived the mass resolutions for Helium obtained with ToF and calorimeter as a function of the rigidity, which is shown in Fig.~\ref{im:he_mass_resolution}, compared with the predictions as obtained and derived from the CERN tests. 
The full black line illustrates the predicted  overall mass resolution for a \hef\ particle resulting from three contributions (shown as dotted lines): rigidity ($MDR_{spec}$), multiple scatter ($MDR_{cou}$), and velocity via ToF (time resolution 100 ps). As one can see the experimental results on the mass resolution obtained from the combination of rigidity and velocity from the ToF system follow nicely the prediction. This gives us confidence that we understand our instrument. It can also be seen that the multiple $dE/dx$ from the calorimeter provide a better mass resolution at higher energies than measuring the velocity via the ToF. This allows us to extend the \pam\ measurements on isotopes to higher energies. It can also be clearly seen that the multiple scatter sets the lower limit of the mass resolution at rigidities below roughly 2 GV no matter how high the MDR of the spectrometer is.

\subsection{Raw isotope numbers}
We had two complementing experimental methods to separate the isotopes: The combination of magnetic spectrometer either with the  ToF or with the multiple $dE/dx$ measurements within the calorimeter. In the following we will describe these procedures separately.

The isotope separation as well as the determination of isotope fluxes was
performed identical to \cite{2013ApJ...770..2} in intervals of kinetic energy per nucleon. Since the magnetic spectrometer measures the rigidity of particles and not the kinetic energy, this means that different rigidity intervals 
have to be analyzed \textbf{depending on the mass of the isotope under study}. For example
Fig.~\ref{im:beta_fit_h} shows the 1/$\beta$ distributions used to select \prot\ 
(top panel) and \deu\ (bottom panel) in the kinetic energy interval
0.361 - 0.395 GeV/n corresponding to 0.90 - 0.95 GV for \prot\ and 1.80 - 1.89
GV for \deu. 

\subsubsection{Raw isotope numbers with the ToF}

\begin{figure}[t]
    \centering
    \plotone{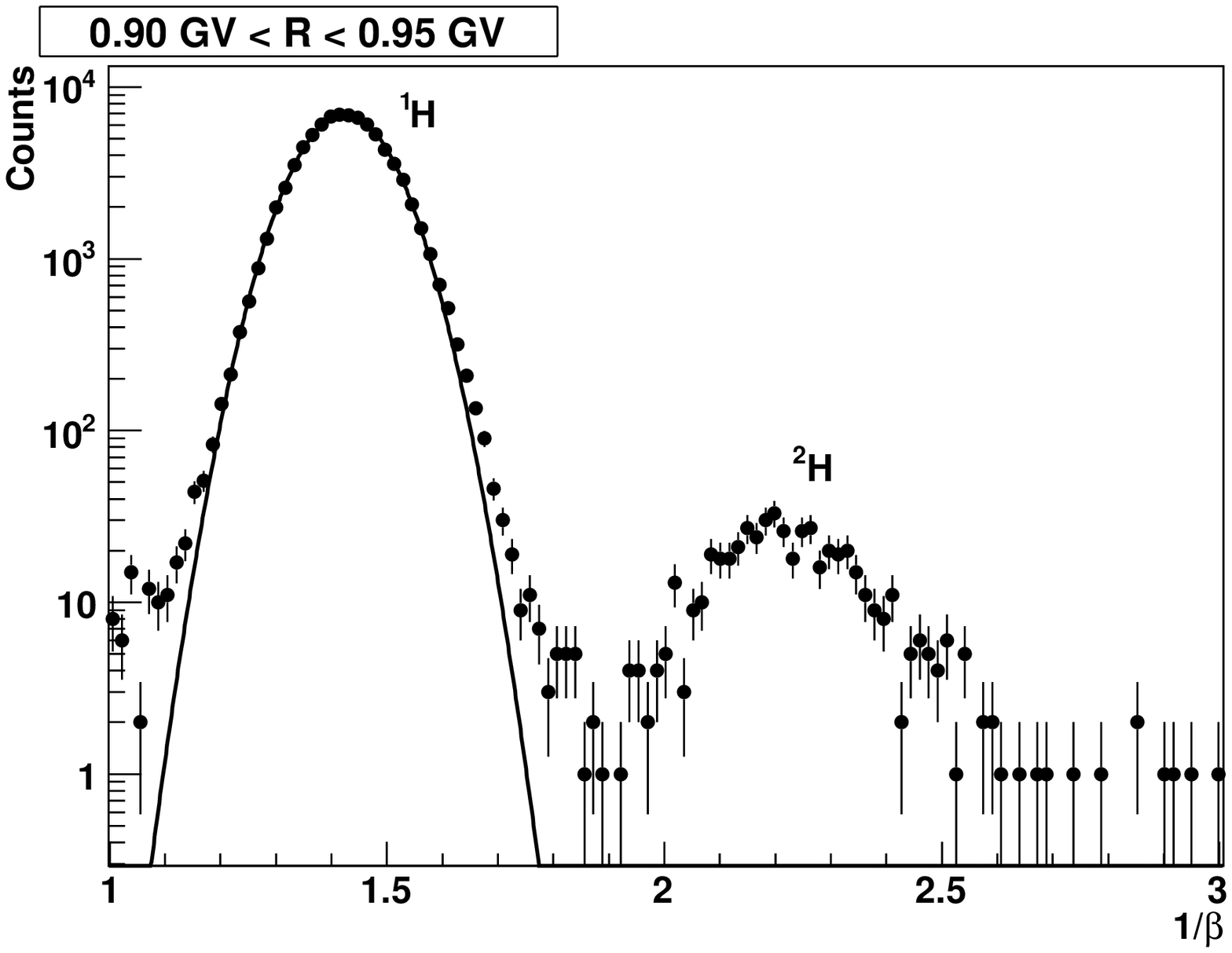}
    \plotone{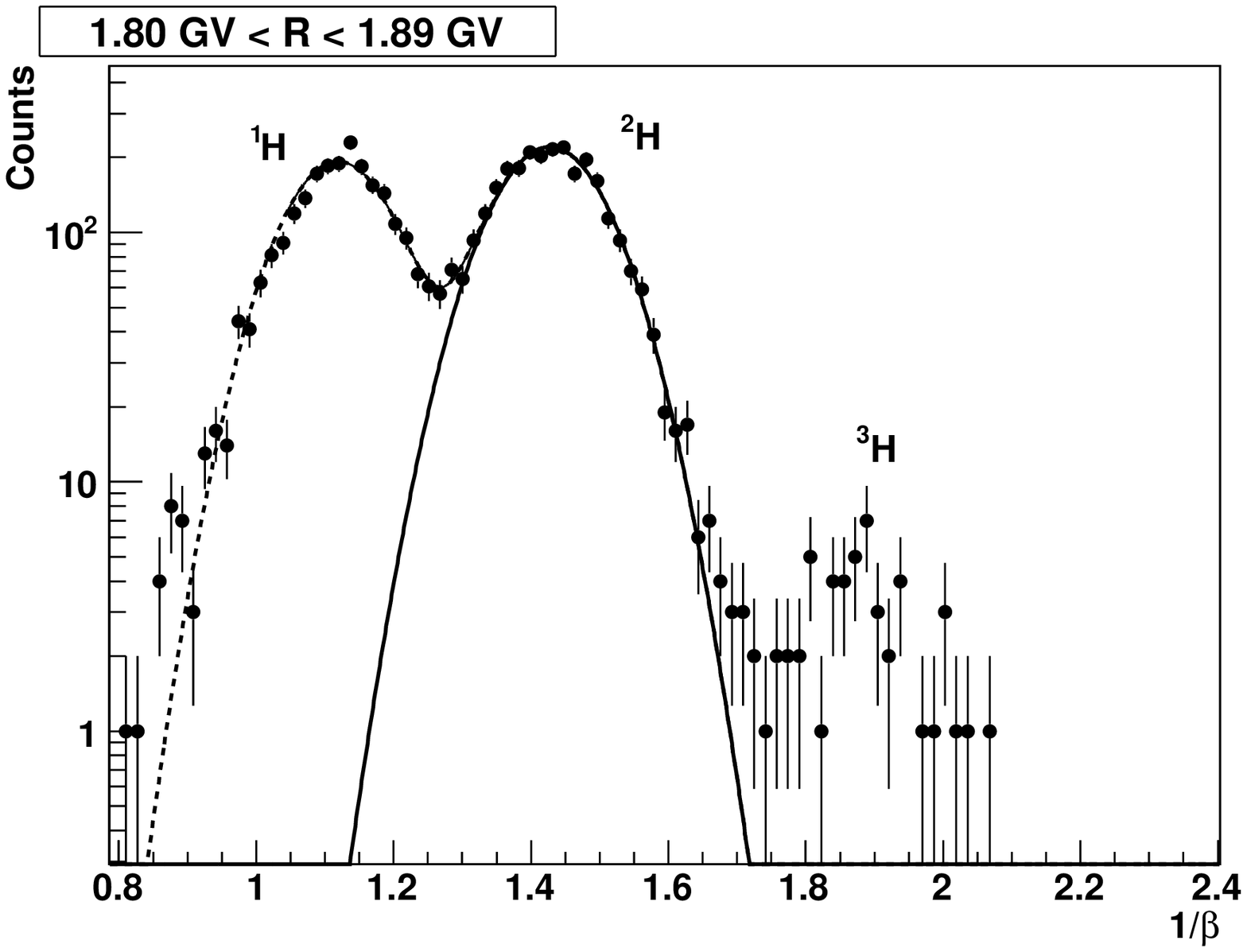}
    \caption{$1/\beta$ distributions for hydrogen in the 0.361 - 0.395 GeV/n kinetic energy range for \prot\ (top) and \deu\ (bottom). The dashed line shows the combined fit (only for \deu) while the solid line shows the \prot\ and \deu\ individual gaussians. Note that the \prot\ component in the \deu\ distribution in the bottom plot is suppressed by the additional selection cuts on the energy loss in ToF and tracker. In the bottom figure the small fraction of \textsuperscript{3}H events is visible. }
    \label{im:beta_fit_h}
\end{figure}

The particle counts in each rigidity range were derived in a similar manner as in \cite{2013ApJ...770..2} by fitting gaussians to the 1/$\beta$ distributions as shown by the solid lines in Fig.~\ref{im:beta_fit_h}.
Instead of mass distributions (shown in Fig.~\ref{im:he_mass_histograms}) 1/$\beta$ distributions were chosen since the shape of a 1/$\beta$ distribution is gaussian, while the mass distribution is not. (Note that for the estimation of the mass resolution in section \ref{sec:mass_resolution} this feature could be neglected). 

\begin{figure}[t]
    \centering
    \plotone{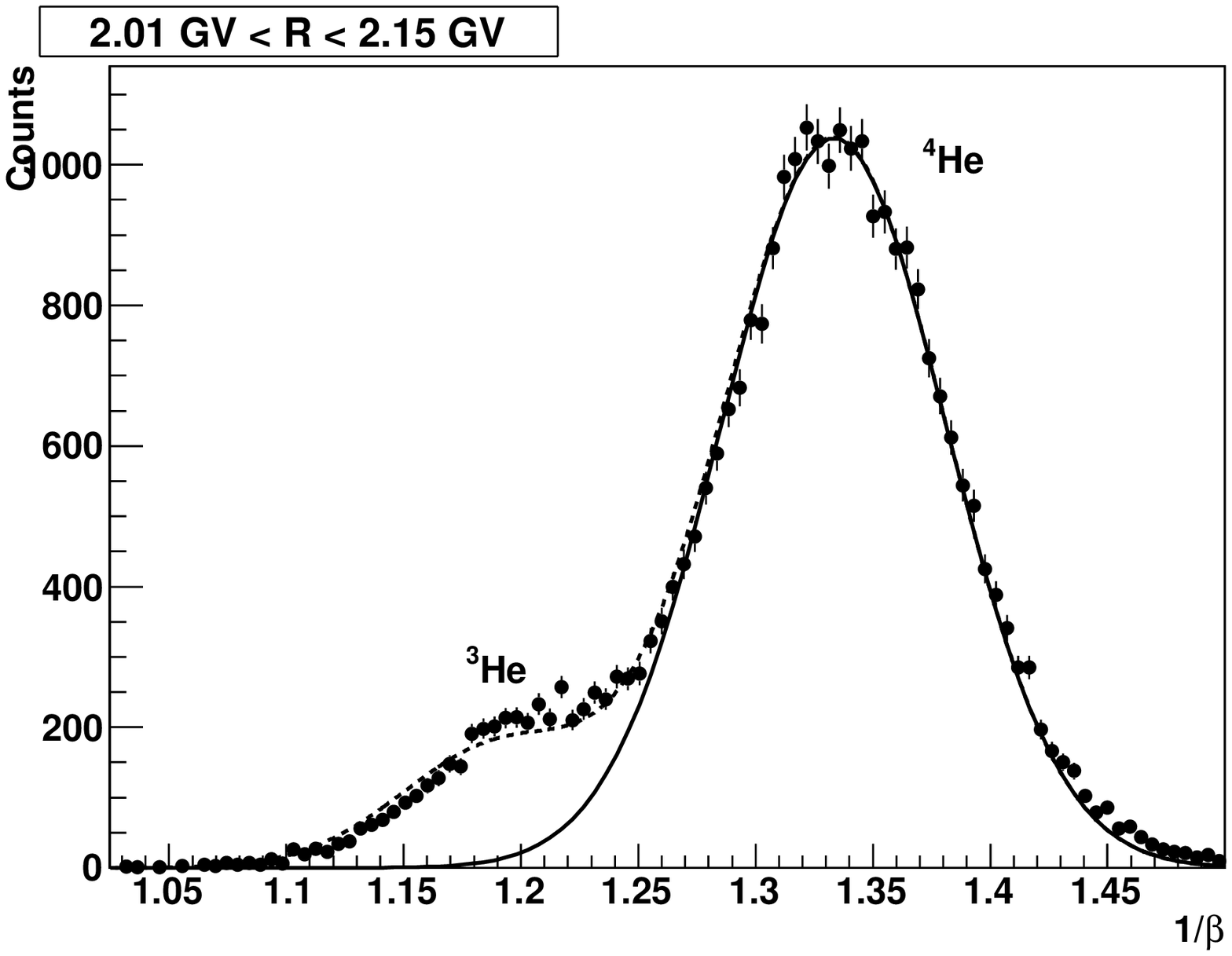}
    \plotone{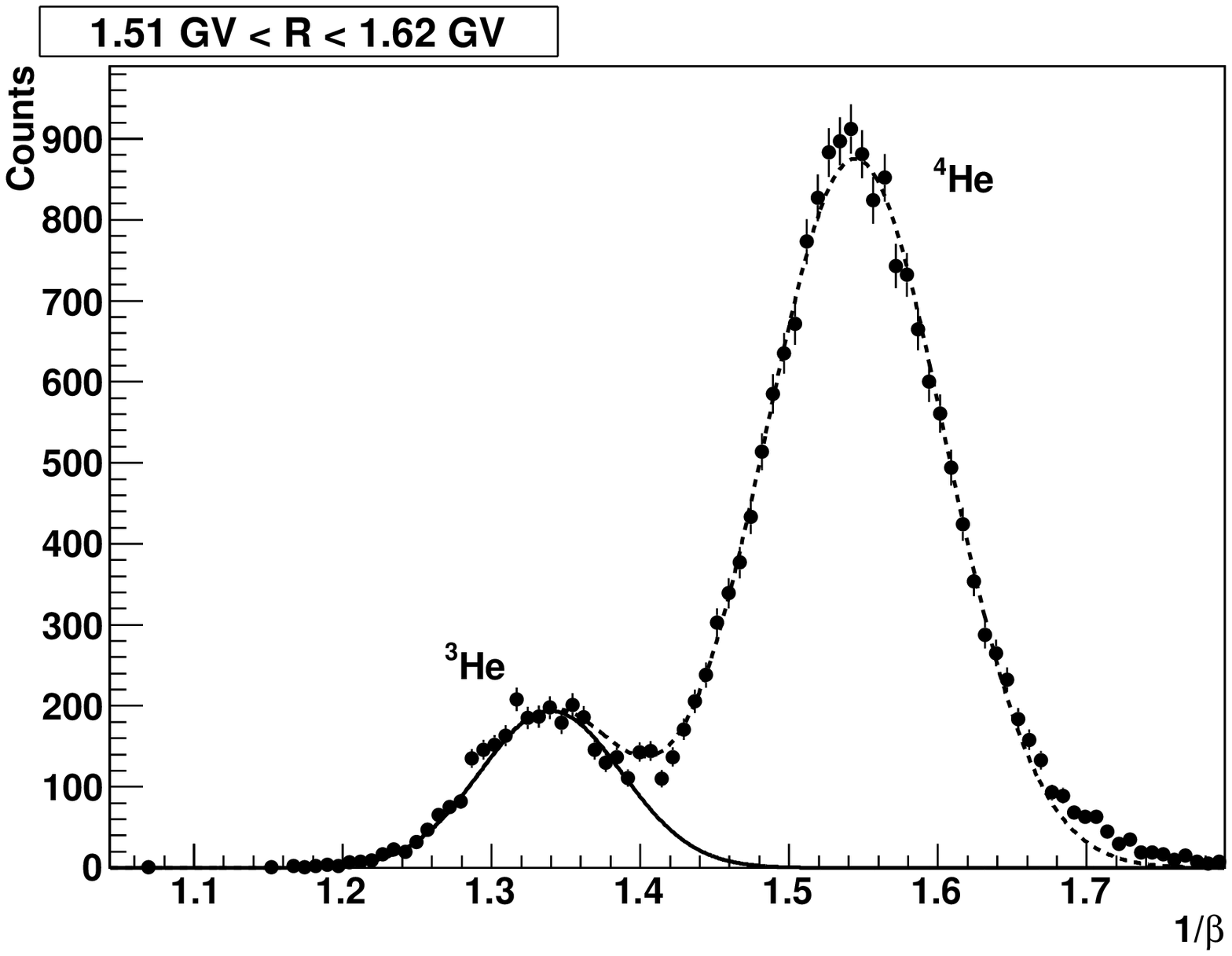}
    \caption{$1/\beta$ distributions for helium in the 0.439 - 0.492 GeV/n kinetic energy range for \hef\ (top) and \het\ (bottom). The dashed line shows the combined fit while the solid line shows the \hef\ and \het\ individual gaussians.}
    \label{im:beta_fit_he}
\end{figure}

The \prot\ peak in Fig.~\ref{im:beta_fit_h} (top) is well pronounced and barely affected by the shape of the neighbouring \deu\ distribution. For that reason a single gaussian was fitted to the \prot\ peak as shown in Fig.~\ref{im:beta_fit_h} (top). For the fitting method the ROOT analysis package \citep{ROOT} was used. 
The gaussian fit to  the \deu, \het\ and \hef\ distributions  becomes a little more complicated since the neighbouring isotopes are quite abundant and have an impact on the fitting. For that reason we applied a suppression procedure to the abundant neighbour which will be more discussed in the following chapter. Consequently  we applied a double gaussian fit to the histograms, see Fig.~\ref{im:beta_fit_h} (bottom) and Fig.~\ref{im:beta_fit_he}, and the whole process was done in a more elaborated manner compared to the analysis presented in \cite{2013ApJ...770..2}.
We went through three steps: We first did the double gaussian fitting with all six parameters left free (mean, sigma and peak of both isotopes) and then analyzed how the values for the means and the sigmas varied with kinetic energy. They followed a trend in kinetic energy and in a second step we fitted appropriate functions sigma=f($E_kin$) and mean=f($E_kin$) to them. All the means and the sigmas followed nicely this trend except at the high energy end, which may be due to the increasing contribution from the more abundant neighbouring isotopes. As a consequence we decided to perform the final fitting to the distributions by fixing the mean and the sigmas according to the function and used in the final double gaussian fitting process only two free parameters: the two heights of the curve's peaks.

Fig.~\ref{im:beta_fit_he} shows the 1/$\beta$ distributions used to select \het\
(bottom panel) and \hef\ (top panel) in the kinetic energy interval
0.439 - 0.492 GeV/n corresponding to 1.51 - 1.62 GV for \het\ and 2.01 - 2.15
GV for \hef.  

\subsubsection{Raw isotope numbers with the ToF: Suppression of abundant \prot\ and \hef}\label{sec:prot_rejection}
As it was already discussed in \cite{2013ApJ...770..2}, the large proton background in the $Z = 1$ sample requires an additional selection in the ToF analysis to suppress the protons at higher energies (roughly 500 MeV/n). Otherwise the gaussian fit for the protons in the $1/\beta$ distributions affects the fit for the much less abundant \deu\ neighbour, especially at higher energies, where the mass resolution is not sufficient for a clear particle separation.
To suppress the abundance of protons, we choose the energy loss measurements in the silicon layers of the tracking system and in the scintillators of the ToF versus the rigidity. The tracking system provides up to 12 energy loss measurements while the ToF provides six. In order to further improve the separation between the isotopes we did not take the mean of the $dE/dx$ measurements but choose the lowest one. This minimizes the Landau fluctuations similar to the truncated mean technique which we use in the calorimeter analysis. 
The two cuts were chosen in such a way that for low energies the protons could be rejected down to a very low level, by keeping practically all \deu. For more details of the specific cuts see Fig. 6 in \cite{2013ApJ...770..2}.

At higher energies the proton contamination will increase plus there will be a loss of \deu. This efficiency was studied with the clean \deu\ sample provided by the calorimeter, and was taken into account in the calculation of the fluxes (section \ref{sec:flux_determination}).

In the $Z = 2$ data the level of the \hef\ background in the \het\ sample is much smaller compared to the $Z = 1$ data, but similar checks like the one described above showed that also in this case a soft cut to suppress \hef\ at higher energies improved the \het\ selection. We used a cut analogue to the $Z = 1$ analysis based on the lowest energy release in the tracking system. Note that this suppression was not used for the $Z = 2$ analysis in \cite{2013ApJ...770..2}.

\subsubsection{Raw isotope numbers with the calorimeter}

For the ToF system the 1/$\beta$ distributions were analyzed by fitting with gaussians. The $dE/dx$ distributions of the calorimeter have a non-gaussian shape, hence one has to model the expected distributions of the observable quantities and then perform likelihood fits.
We used the ``RooFit'' toolkit \citep{2003physics...6116V} for the likelihood fits. First one has to create the expected $dE/dx$ distributions (``probability density function'': PDF) for each isotope. We used the full Monte Carlo simulation of the \pam\ apparatus based on the \texttt{GEANT4} code \citep{Geant4}, which has been already described in \cite{2013ApJ...770..2}, for this task.

When taking the simulated energy loss in each layer as coming from \texttt{GEANT4}, we noticed that the resulting PDFs showed a slight mismatch from the flight data. We found that the width of the histograms was smaller than in the real data, also there was a small offset of about 1 - 2\%. We applied a multiplicative factor to the simulated energy loss in a layer, plus adding a gaussian spread of the signal of a few percent.

\begin{figure}[t]
    \centering
    \plotone{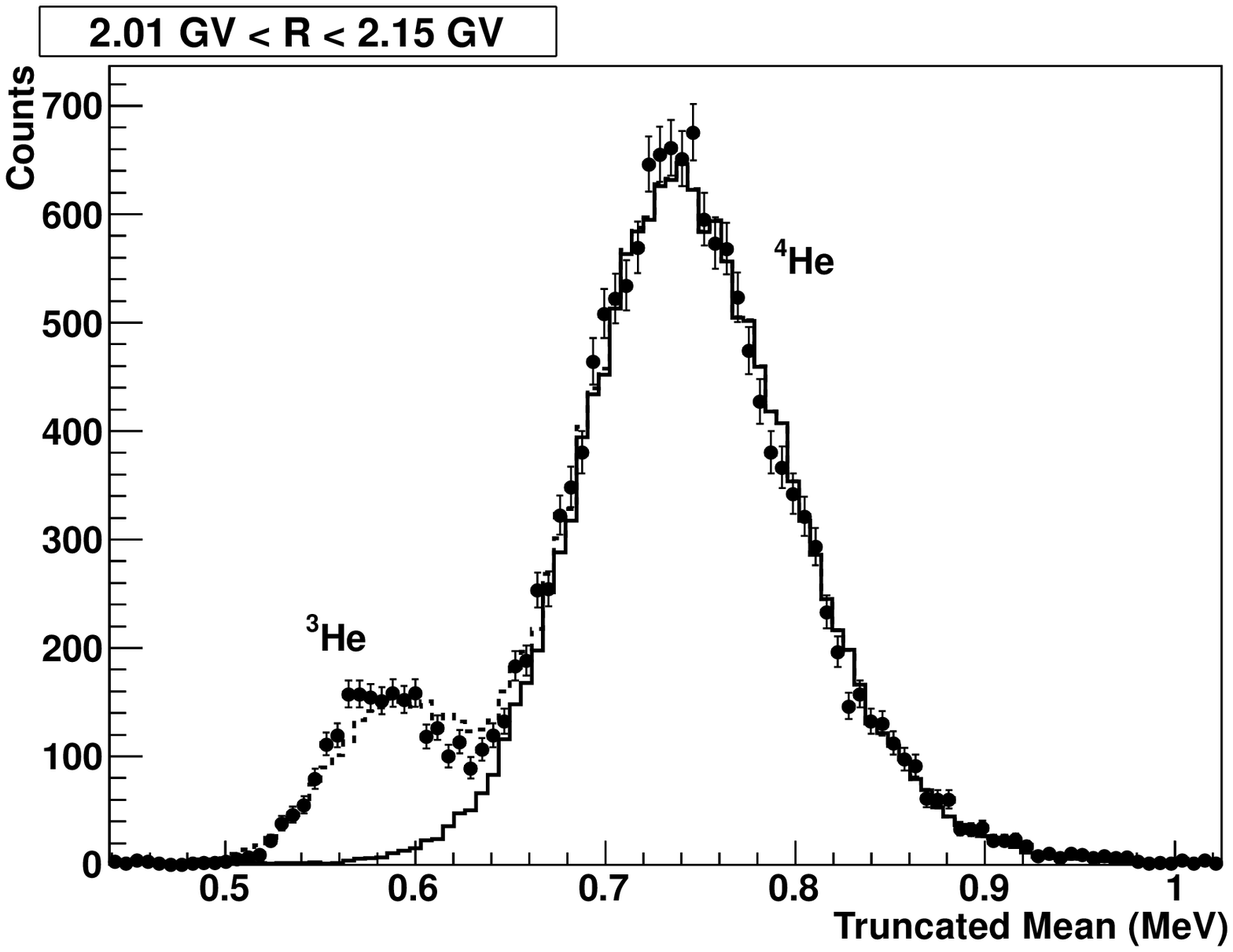}
    \plotone{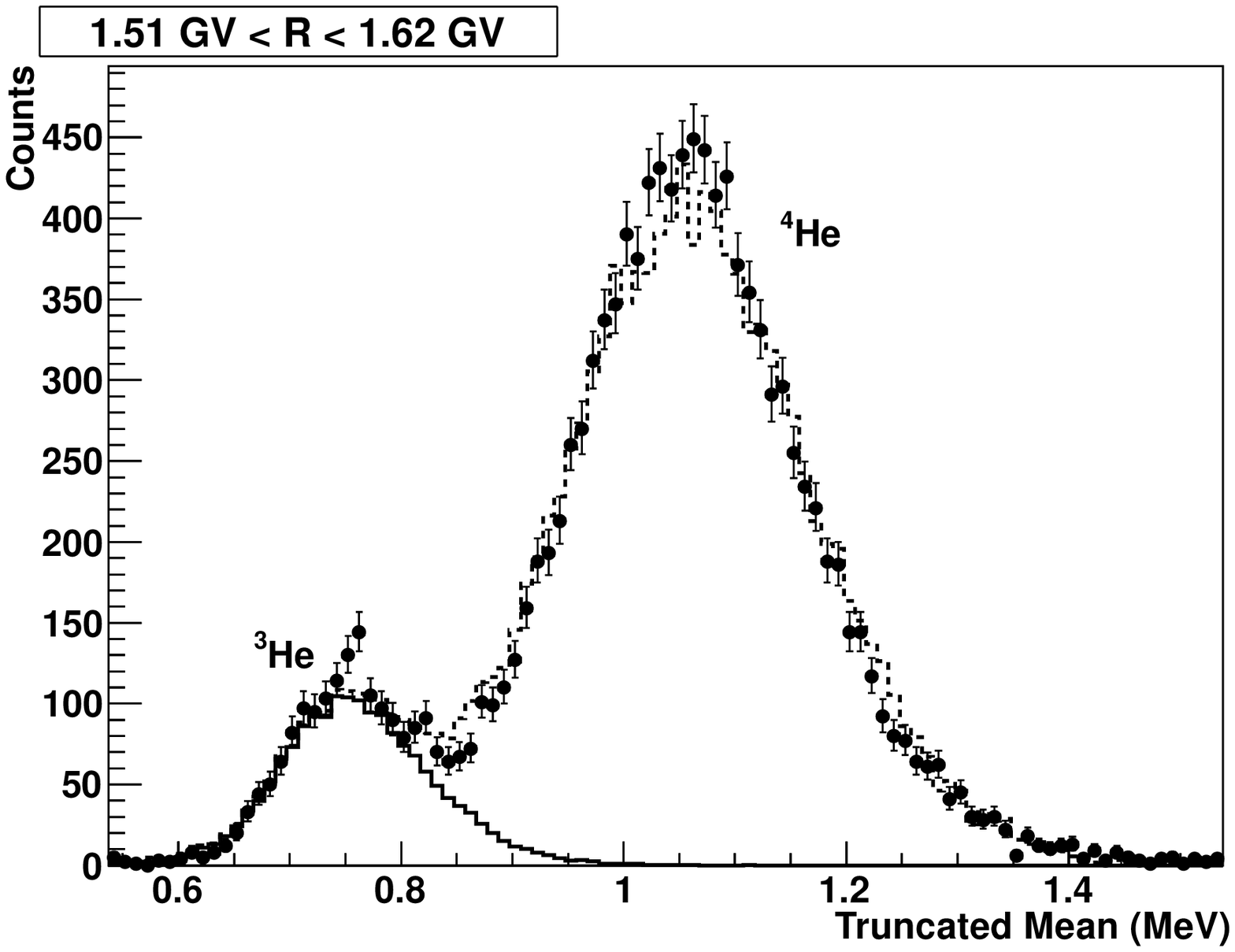}
    \caption{
Example of the truncated mean distributions for helium in the 0.439 - 0.492 GeV/n kinetic energy range for \hef\ (top) and \het\ (bottom). The dashed line shows how the combined fit using the two PDFs derived with the modified \texttt{GEANT4} simulation matches the data points  (black points), while the solid line shows the estimated \hef\ and \het\ individual signals.
}
    \label{im:fig_helium2}
\end{figure}

As an example we show in Fig.~\ref{im:fig_helium2} the truncated mean distributions for helium in the 0.439 - 0.492 GeV/n kinetic energy range for \hef\ (top) and \het\ (bottom).
The dashed line shows how the combined fit using the two PDFs derived with the modified \texttt{GEANT4} simulation matches the data points (black points) while the solid line shows the estimated individual \het\ and \hef\ signals. 
The kinetic energy range is the same as shown in Fig.~\ref{im:beta_fit_he} for the ToF, the difference in the isotopic separation is clearly visible.

Due to the redundant detectors of \pam\ we were able to test the simulated PDFs with real data from the instrument. Fig. \ref{im:dedx} illustrated the mass resolution which can be obtained by combining the mean $dE/dx$ measurement in the tracker and the rigidity measurement with the magnetic spectrometer, while Fig. \ref{im:beta_r} showed a similar picture using the velocity measurement from the ToF and the rigidity measurement.

\begin{figure}[t]
    \centering
    \plottwo{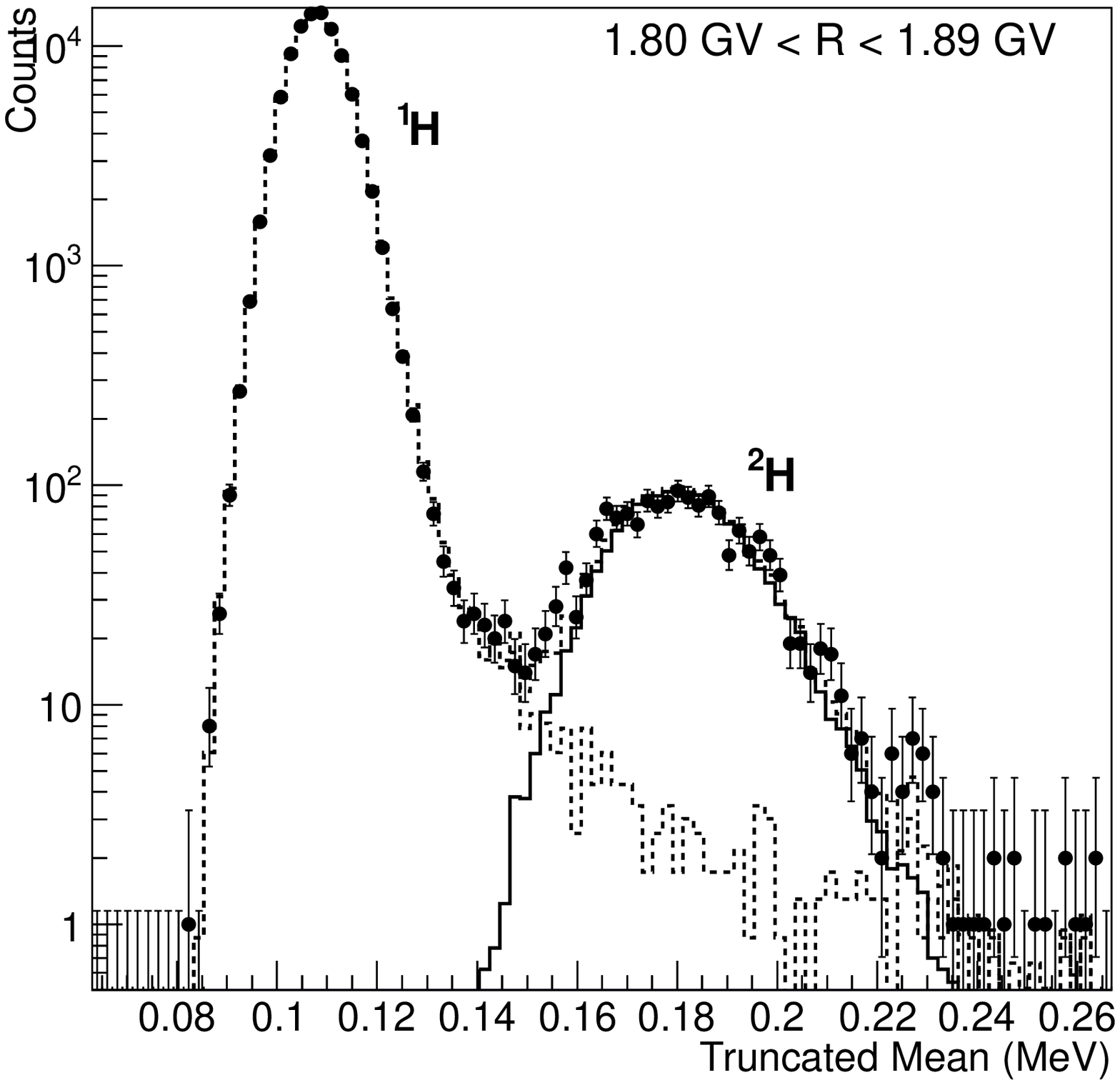}{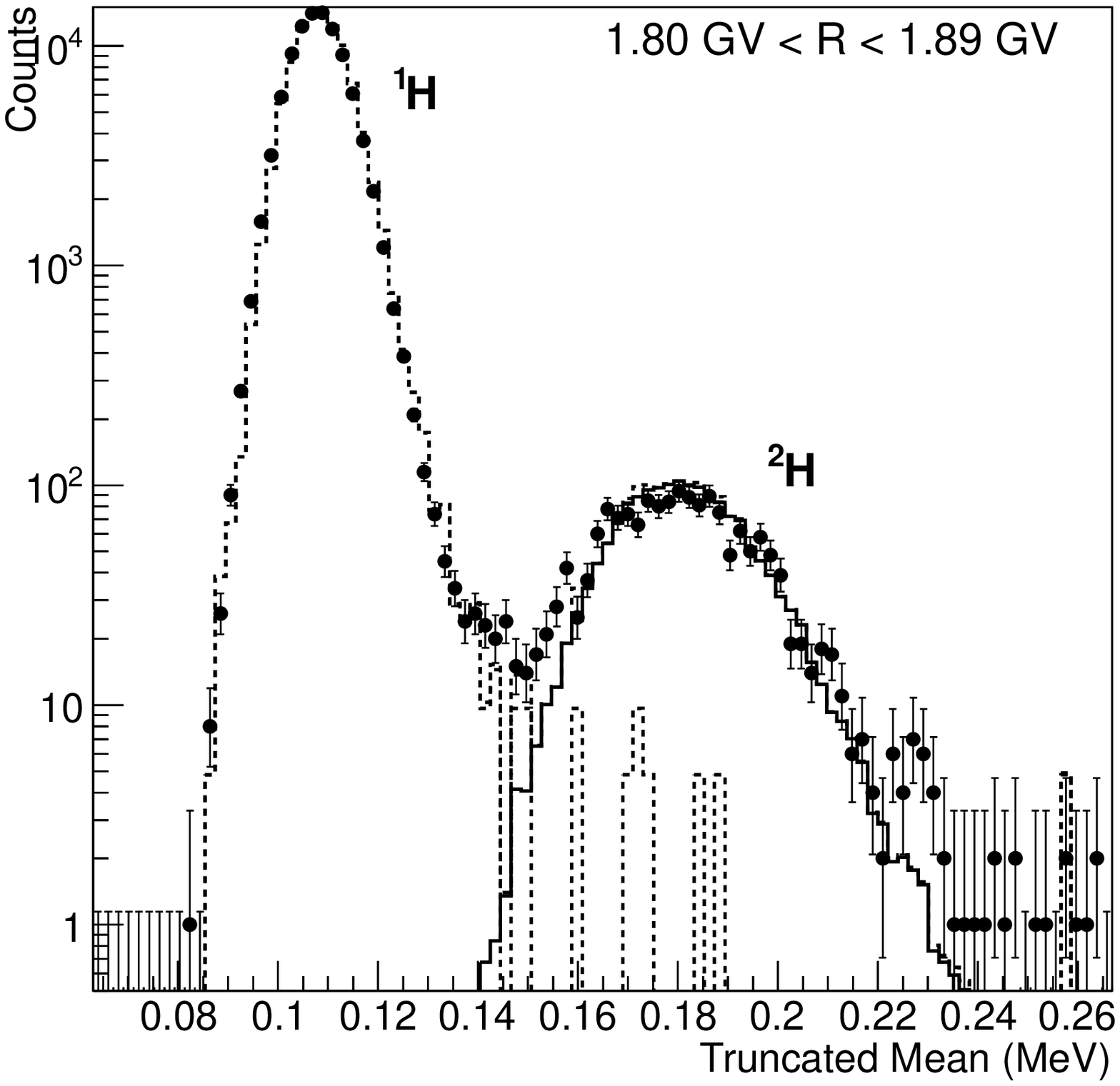}
    \caption{Example RooFit of \deu. Black points: data, dashed line: \prot\ model, solid line: \deu\ model. The \deu\ model is taken from the simulation for both plots, while in the left plot the \prot\ model is derived from flight data and from simulation in the right plot.}
    \label{im:h1_flight_simu}
\end{figure}

By using appropriate selection cuts we separated a proton sample and asked for the energy loss response in the calorimeter. This result was then compared to the simulated distribution. Such a comparison is shown in Fig. \ref{im:h1_flight_simu} for the for the rigidity interval 1.80 GV - 1.89 GV. 
(The rigidity interval in this figure is the same as in Fig.~\ref{im:beta_fit_h} for the ToF, also here the difference in the isotopic separation is clearly visible).

It can clearly be seen that the flight data proton PDF shows a larger tail into the \deu\ histogram compared to the simulated PDF. This tail will cause to derive a lower number of \deu\ counts compared to the simulated PDF (the difference in the \deu\ counts is about 8\% in this example). 

While the creation of PDFs from flight data for \deu, \het\ and \hef\ is restricted to lower energies due to the limited isotopic separation, we created PDFs for protons up to 4 GV using strict selection cuts. In principle a clear separation between \prot\ and \deu\ is not possible at these rigidities, but since the \prot\ are so dominant, the contamination of \deu\ in the \prot\ sample should be very small. We observed that the tail in Fig. \ref{im:h1_flight_simu} is only visible for medium energies, where our selection cuts for the truncated mean are quite soft. Since a small number of layers are used to derive the truncated mean, fluctuations are probably still significant, and it seems that the \texttt{GEANT4} simulation cannot fully reproduce the actual energy loss under these circumstances.

At higher energies, where our selection cuts for the calorimeter are stricter, the tails in the flight data PDF disappeared, resulting in a good agreement with the simulated PDF again. We decided to take simulated PDFs except for the \prot\ model for the ``RooFit'' analysis in this paper.

One could argue that also for the other isotopes it might be that the simulated PDFs do not show the correct shape, missing the tail which is visible for \prot, so for example we might underestimate the number of \het\ which contribute to the \hef\ distribution. However, we found that the maximum difference in the \deu\ counts at medium energies was at most 10\%, with the number of \prot\ exceeding the number of \deu\ in the distribution by a factor of 50 - 100. In comparison the \het\ / \hef\ ratio is around 0.2, so one can expect that the effect of missing tails in the simulated PDFs will have a negligible influence to the \het\ and \hef\ counts.

\subsection{Flux Determination}\label{sec:flux_determination}

\begin{figure}[t]
    \centering
    \plotone{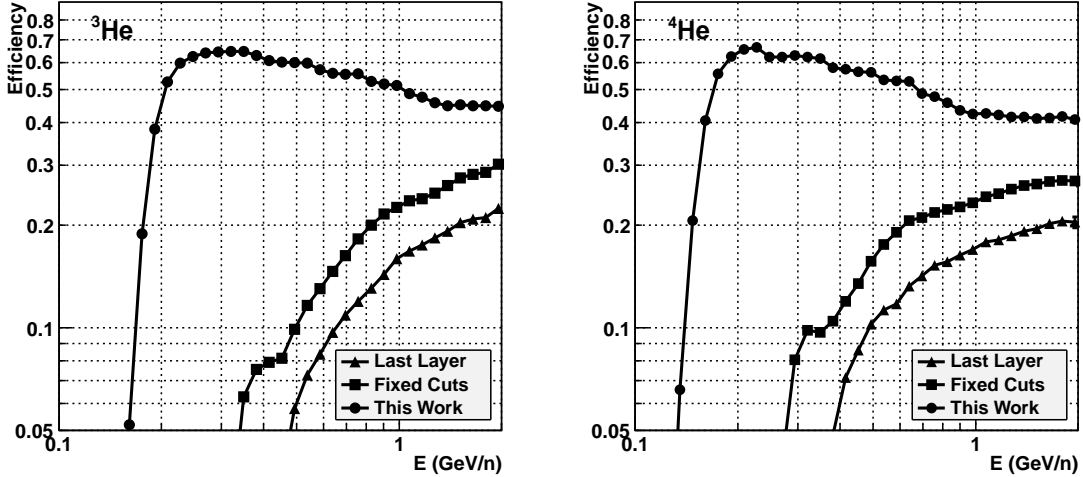}
    \caption{Calorimeter selection efficiency for helium derived with simulated data: triangles: $dE/dx$ signal in the last layer, squares: fixed cuts (30 hit silicon strips), circles: dynamic cuts used for this work. }
    \label{im:he_calo_effi1}
\end{figure}

To derive the isotope fluxes, the number of \prot\ and \deu\ events in the $Z=1$ sample and the number of \het\ and \hef\ events in the $Z=2$ sample had to be corrected for the selections efficiencies, particle losses, contamination and energy losses. 
Most of the corrections could be directly taken from \cite{2013ApJ...770..2}, only some efficiencies in the ToF analysis were changed, for example the efficiency for the suppression of the abundant \prot\ and \hef\ (see section \ref{sec:prot_rejection}).
A new correction is the efficiency for the calorimeter, which is shown in 
Fig.~\ref{im:he_calo_effi1} for specific selection cuts. The selection cuts for the actual analysis were described in section \ref{sec:calo_principle}. One can nicely see that this approach gives a quite high efficiency showing a rather constant behaviour down to about 200 MeV/n where the efficiency then shows a steep decrease.
As mentioned already above one can do these selections within the calorimeter analysis in various ways, but the efficiencies are very sensitive to the applied cuts. This is illustrated in Fig.~\ref{im:he_calo_effi1} for two other cut conditions: A strict cut where the particle has to fully traverse the calorimeter and produce a signal in the last layer, and a more relaxed cut (30 hit silicon strips). As one can see,  the efficiencies of these cuts are quite low and show a steep drop already at 300 - 400 MeV/n.  Preliminary results for the relaxed cut have been presented in earlier publications (\cite{2013JPhCS.409a2030M},
\cite{2013ICRC...0233}).

\begin{figure}[t]
    \centering
    \plotone{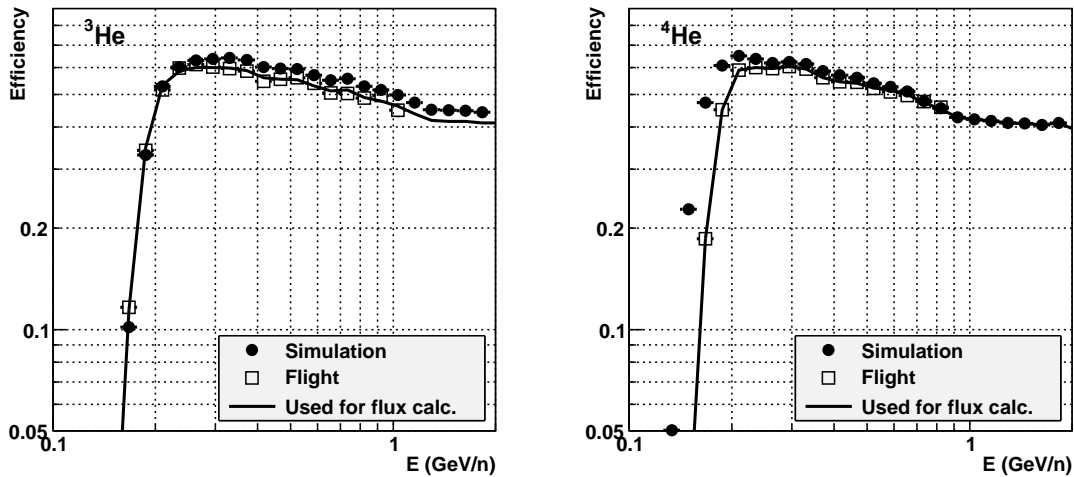}
    \caption{Comparison between the efficiency derived with simulated data (circles) and flight data (squares). The solid line shows the efficiency used for the analysis.}
    \label{im:he_calo_effi2}
\end{figure}

The comparison between efficiencies derived with simulated data and the ones derived with flight data (using ToF and tracker $dE/dx$ for selecetion) is shown for $Z = 2$ particles in Fig.~\ref{im:he_calo_effi2}.

As one can see, there is a good agreement at low energies for \het\ while there is some difference at higher energies, while it is the opposite for \hef. We decided to use the flight data effiencies at lower energies and then for higher energies (roughly around 700 MeV/n in Fig.~\ref{im:he_calo_effi2}), follow the trend of the simulated efficiencies by applying a constant correction factor. The same method was used for \deu\ efficiency, while for \prot\ we used the flight data efficiency for the full energy range.

The following corrections are taken from \cite{2013ApJ...770..2} without changes and we refer to this paper for more details:
\begin{itemize}
\item 
Due to hadronic interactions in the aluminum pressurized container (2 mm thick) and the top scintillators helium and hydrogen nuclei might be lost. The correction factor $b(E)$ is different for each isotope and has been derived from the Monte Carlo simulation, being $\simeq 6\%$ for \prot, $\simeq 10\%$ for \deu, and $\simeq 13\%$ for both helium isotopes,.
\item  The nominal geometrical factor $G_F$ of \pam\ is almost constant above 1 GV, with the requirements on the fiducial volume corresponding to a value of $G_F = 19.9$  cm\textsuperscript{2} sr, for lower energies the bending of the particles track leads to a decrease. The nominal geometrical factor $G_F$ was multiplied with the correction factor $b(E)$ to get an effective geometrical factor $G(E)$  see Fig. 7 in \cite{2013ApJ...770..2}.
\item Contribution to \deu\ from inelastic scattering of \hef: This background was derived from the simulation and subtracted from the raw \deu\ counts. The contamination is in the order of $\simeq 10\%$ at 100 MeV/n, going down with increasing enery ($\simeq 1\%$ at 600 MeV/n), see Fig. 8 in \cite{2013ApJ...770..2}. The contamination in the \het\ sample from \hef\ fragmentation was also evaluated and was found to be very small (less than $1\%$), this was included in the systematic uncertainty of the measurement. 
\item The measured particle spectra are distorted due to particle slowdown (caused by the energy loss) and the finite resolution of the spectrometer. We used a Bayesian unfolding procedure \citep{dagostini} to derive the number of events at the top of the payload (see \cite{2011Sci...332...69A}).
\end {itemize}

The differential flux is then given by
\begin{equation}
  \Phi_{\text{ToP}} (E) = \frac{N_\text{ToP}(E)}{T G(E) \Delta E}
\end{equation}
where $N_\text{ToP}(E)$ is the unfolded particle count (corrected for the selection efficiencies) for energy $E$, ,
$\Delta E$ is the energy bin width, and $G(E)$ is the effective geometrical factor as described above. 
The live time, $T$, depends on the orbital selection as described in section \ref{sec:event_selection} and is evaluated by the trigger system \citep{bru08}.

\subsection{Systematic uncertainties}\label{sec:systematics}

The systematic uncertainties presented in \cite{2013ApJ...770..2} have been reviewed and updated to the new analysis methods when neccessary.

The event selection criteria described in section \ref{sec:event_selection} were similar to previous work on high energy proton and helium fluxes, see \cite{2011Sci...332...69A}. In that paper the systematic errors of the selection have been studied using flight data and simulations, resulting in a quoted systematic uncertainty of ca. $4\%$. This error is used also in this work.

As already decribed in section \ref{sec:prot_rejection}, the quality of the Gaussian fit procedure in the ToF analysis was tested using the truncated mean of the energy deposited in the electromagnetic calorimeter to select pure samples of \prot, \deu, \het, and \hef\ from non-interacting events. 
For the abundant particles \prot\ and \hef\ the number of reconstructed events from the Gaussian fit was found to agree with the number of events selected with the calorimeter practically over the full energy range, while for \deu\ and \het\ there were some systematic differences of some percent in the highest energy bins. Note that without the additional cuts which reject the more abundant \prot\ and \hef\ the differences would be much larger. We assigned a systematic uncertainty of $0.5\%$ for low and medium energies increasing 
to $4\%$ at 600 MeV/n for \deu\ and to $3\%$ at 800 MeV/n for \het. For \prot\ the systematic uncertainty was set constant to $0.5\%$ while for \hef\ it was set energy dependent, increasing to $1.5\%$ at 800 MeV/n.

A similar systematic error is assigned to the fit procedure made with the calorimeter. Here deviations between the model PDF and the flight data will transform to a systematic difference in the number of reconstructed events. However, we have no other detector to select pure samples of the isotopes. Therefore we studied how a misplacement of the model PDFs transferred to different particle counts. Similar to the results for the ToF it was found that the effect on the number of reconstructed events was much more pronounced for \deu\ and \het\ compared to \prot\ and \hef, and that the effect increased with energy.
However, the misplacement of the model PDFs can be checked by using the abundant \prot\ and \hef\ as a reference (for example, comparing the peaks in the distributions), thus limiting the systematic differences in the number of reconstructed events to some percent in the highest energy bins. Similar to the systematic error for the ToF, we assigned a systematic uncertainty of $0.5\%$ for low and medium energies increasing to $4\%$ at 1000 MeV/n for \deu\ and to $3\%$ at 1400 MeV/n for \het. The \prot\ the systematic uncertainty  was set constant to $0.5\%$ while for \hef\ it was set again energy dependent, increasing to $1.5\%$ at 1400 MeV/n.

The efficiency of the calorimeter selection was derived using simulated and flight data, as shown for $Z = 2$ particles in Fig.~\ref{im:he_calo_effi2}. For $Z = 2$ data the agreement between the two methods is quite good, and we assigned a conservative systematic error 
of $2\%$ for \hef\ and $3\%$ for \het\ independent from the energy.
For $Z = 1$ particles the difference between the two methods is larger. As stated above, for \prot\ we used the flight data efficiency for the full energy range, which should result in a small systematic error, since the \prot\ are so abundant and therefore the contamination of other particles is negligible. We assigned a conservative systematic error of $2\%$. For the \deu\ efficiency we used the flight data efficiencies at lower energies, but followed the trend of the simulated efficiencies by applying a constant correction factor for higher energies, we estimated a systematic error of $5\%$.

The following systematic uncertainties are taken from \cite{2013ApJ...770..2} without changes and we refer to this paper for more details:
\begin{itemize}
\item
The systematic uncertainty on the \deu\ flux resulting from the subtraction of secondary \deu\ from \hef\ spallation is $1.9\%$ at low energy dropping below $0.1\%$ at 300 MeV/n due to the finite size of the Monte Carlo sample. The validity of the Monte Carlo simulation has been tested in  \cite{2013ApJ...770..2} using the \textsuperscript{3}H component in the flight data sample, see Fig. \ref{im:beta_fit_h}. 
    
\item The systematic uncertainty on the unfolding procedure has been discussed in \cite{2011Sci...332...69A} and was found to be 2\%, independent of energy.
 
\item The selection of galactic particles was described in section \ref{sec:event_selection}, the correction for particles lost due to this selection has an uncertainty due to the size of the Monte Carlo sample. The systematic error decreases from $6\%$ at 120 MeV/n to $0.06\%$ at 1000 MeV/n.

\item The uncertainty on the effective geometrical factor as estimated from the Monte Carlo simulation is $0.18\%$, practically independent of energy.
\end{itemize}  

The systematic uncertainties are included in Tables \ref{tab:hydrogen_tof}, \ref{tab:hydrogen_calo}, \ref{tab:helium_tof}, and \ref{tab:helium_calo} and in Figs. \ref{im:fig_data_h1}, \ref{im:fig_data_he1}, \ref{im:fig_data_h2}, \ref{im:fig_data_he2} and \ref{im:ratios}.

\section{Results and discussion}

\begin{figure*}[t]
    \centering
    \epsscale{0.75}
    \plotone{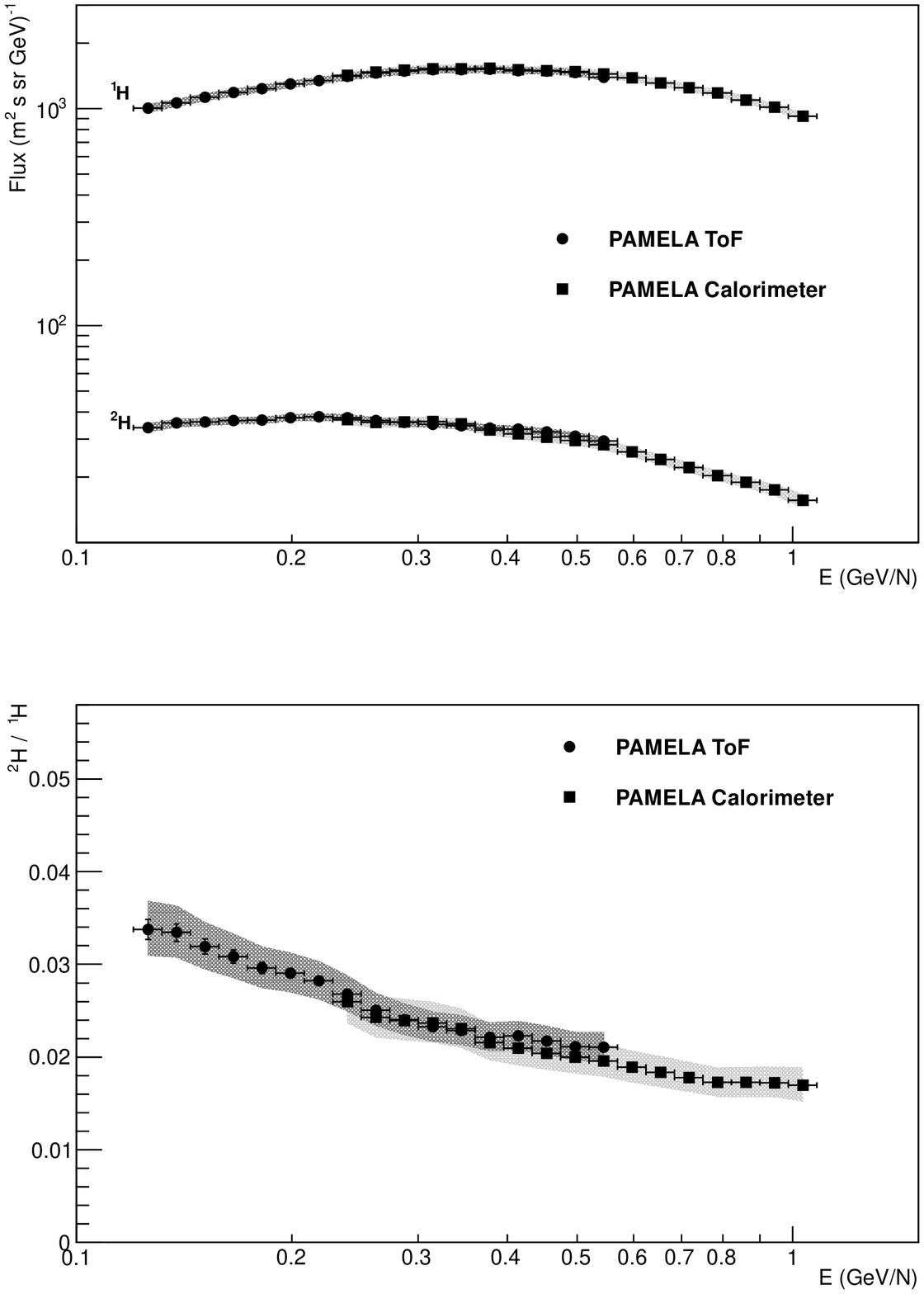}
    \caption{\prot\ and \deu\ absolute fluxes (top) and their ratio (bottom) derived with the ToF (circles) or the calorimeter (squares).
      Error bars show the statistical uncertainty while shaded areas show the systematic uncertainty. }
    \label{im:fig_data_h1}
\end{figure*}

\begin{figure*}[t]
    \centering
    \epsscale{0.75}
    \plotone{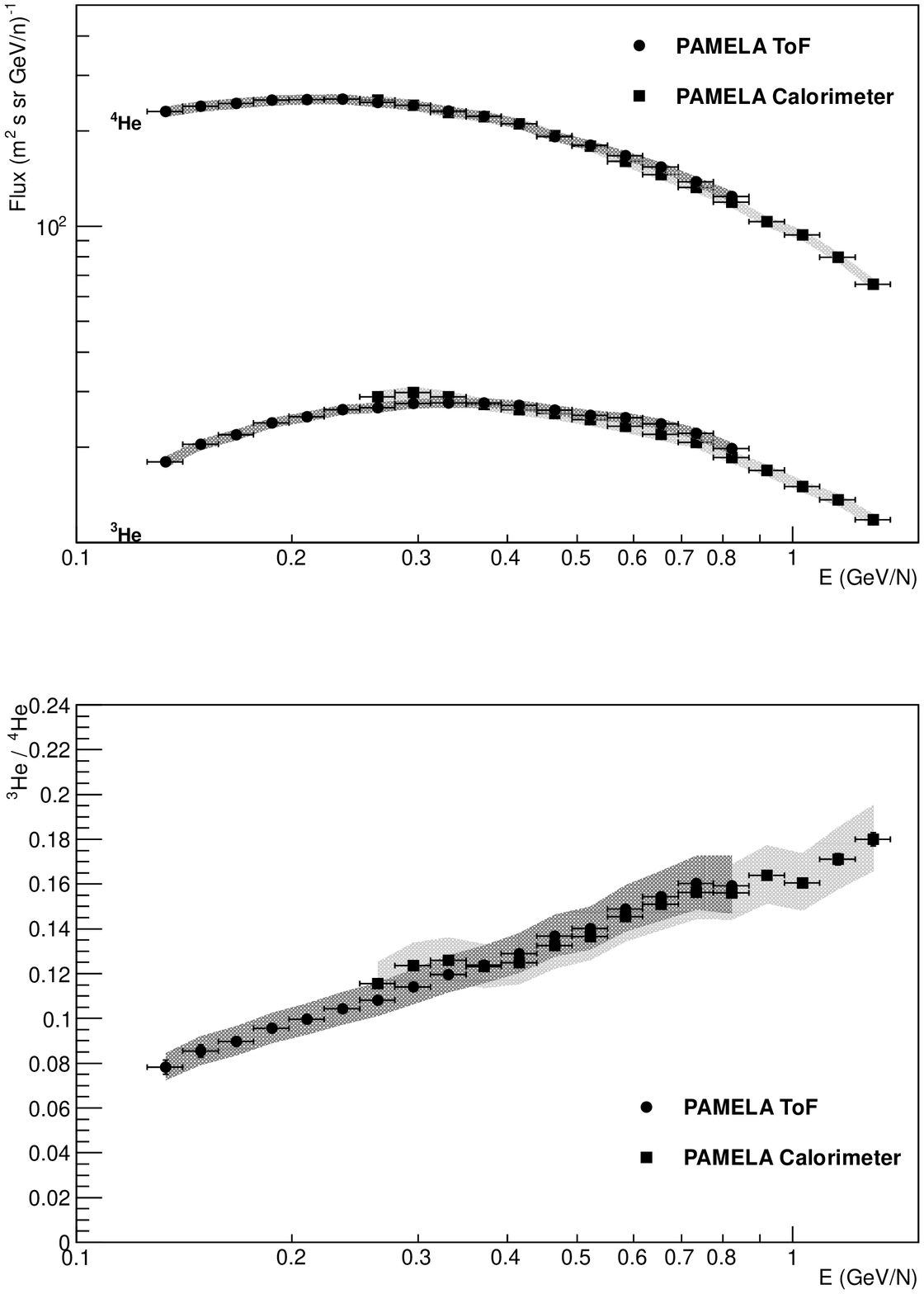}
    \caption{\hef\ and \het\ absolute fluxes (top) and their ratio (bottom) derived with the ToF (circles) or the calorimeter (squares). Error bars show statistical uncertainty while shaded areas show systematic uncertainty.}
    \label{im:fig_data_he1}
\end{figure*}

In Figure~\ref{im:fig_data_h1} and \ref{im:fig_data_he1} we show the hydrogen and helium isotope fluxes (top) and the ratios of the fluxes (bottom) measured with the ToF or the calorimeter. Results are also reported in Tables 
\ref{tab:hydrogen_tof}, \ref{tab:hydrogen_calo}, \ref{tab:helium_tof},  and \ref{tab:helium_calo}. 

It is worth noting that the \pam\ results obtained via the ToF analysis and via the multiple $dE/dx$ measurements with the calorimeter agree very well within their systematic errors. This gives confidence to the results.

In direct comparison with our first paper \citep{2013ApJ...770..2} the results obtained via the ToF analysis in this work show some differences to our earlier results. 
While the \prot\ fluxes show only minor differences, the \deu\ flux is roughly about 5\% higher in this work.
The \hef\ flux is almost 10\% higher at the lowest energy bin, at the highest energies the new \hef\ flux is about 10\% lower, while at medium energies around 400 MeV/n the two results agree.
The new \het\ flux is about 3-4\% lower for most of the energy range, for energies above 500 MeV/n the difference increases and reaches about 15\% for the highest energy bins.
We attribute this to the changes in the fitting procedure (for example, the double gaussian fit for \hef, also the fixing of parameters) and improvements in the efficiency calculation compared to the first paper. 
Based on this more comprehensive analysis presented here these results supersede the previous ones.

\begin{figure*}[t]
    \centering
    \epsscale{0.75}
    \plotone{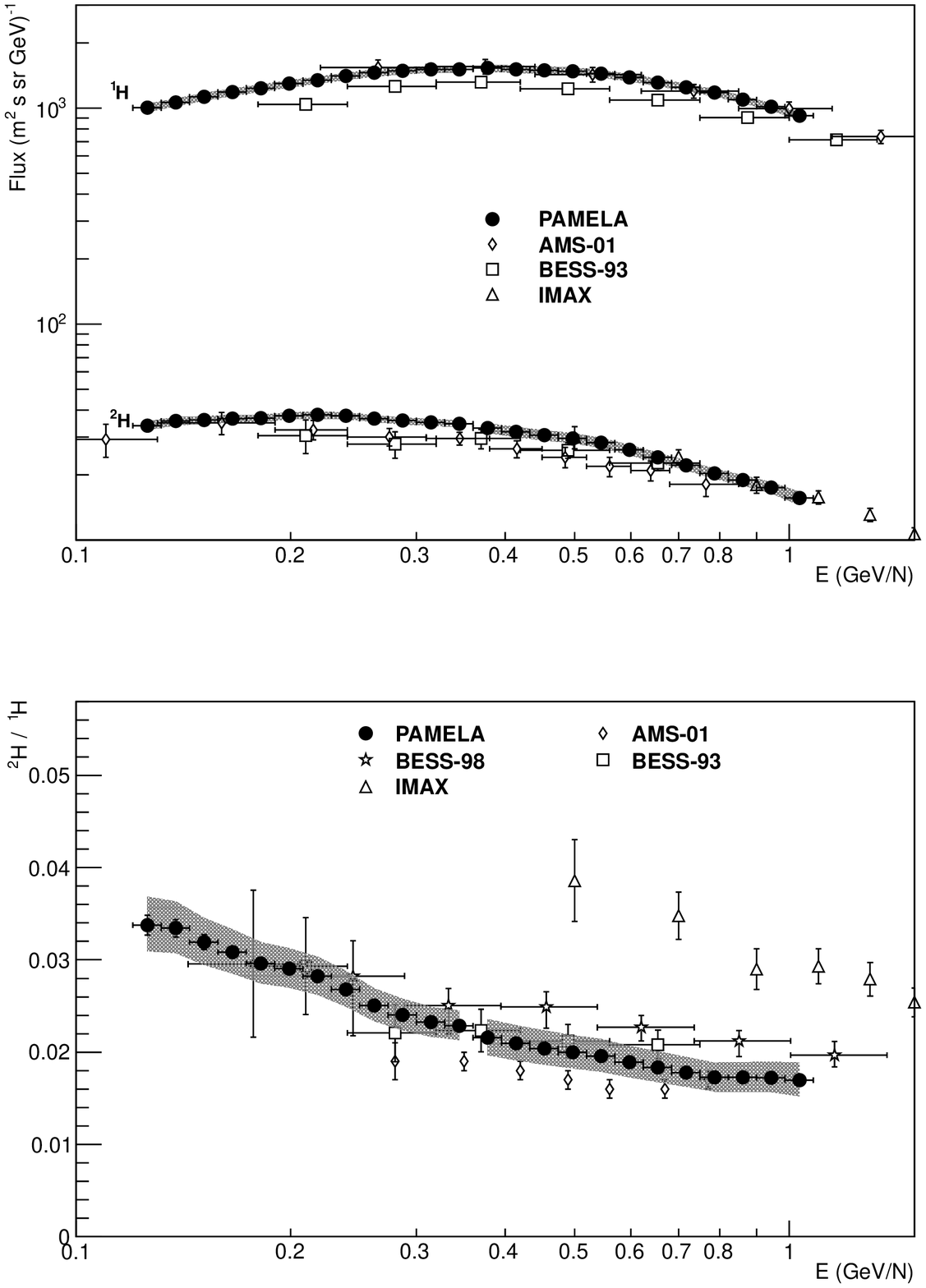}
    \caption{\prot\ and \deu\ absolute fluxes (top) and their ratio (bottom). For energies less than 361 MeV/n the ToF results (Table~\ref{tab:hydrogen_tof}) were used, for higher energies the calorimeter results (Table~\ref{tab:hydrogen_calo}). 
The previous experiments are:  AMS-01 \citep{2002PhR...366..331A,2011ApJ...736..105A_red,2001ICRC...1617}, BESS-93 \citep{2002ApJ...564..244W}, BESS-98 \citep{2005ASR...35..151},   IMAX \citep{2000AIP...528..425}. 
      Error bars show the statistical uncertainty while shaded areas show the systematic uncertainty. }
    \label{im:fig_data_h2}
\end{figure*}

\begin{figure*}[t]
    \centering
    \epsscale{0.75}
    \plotone{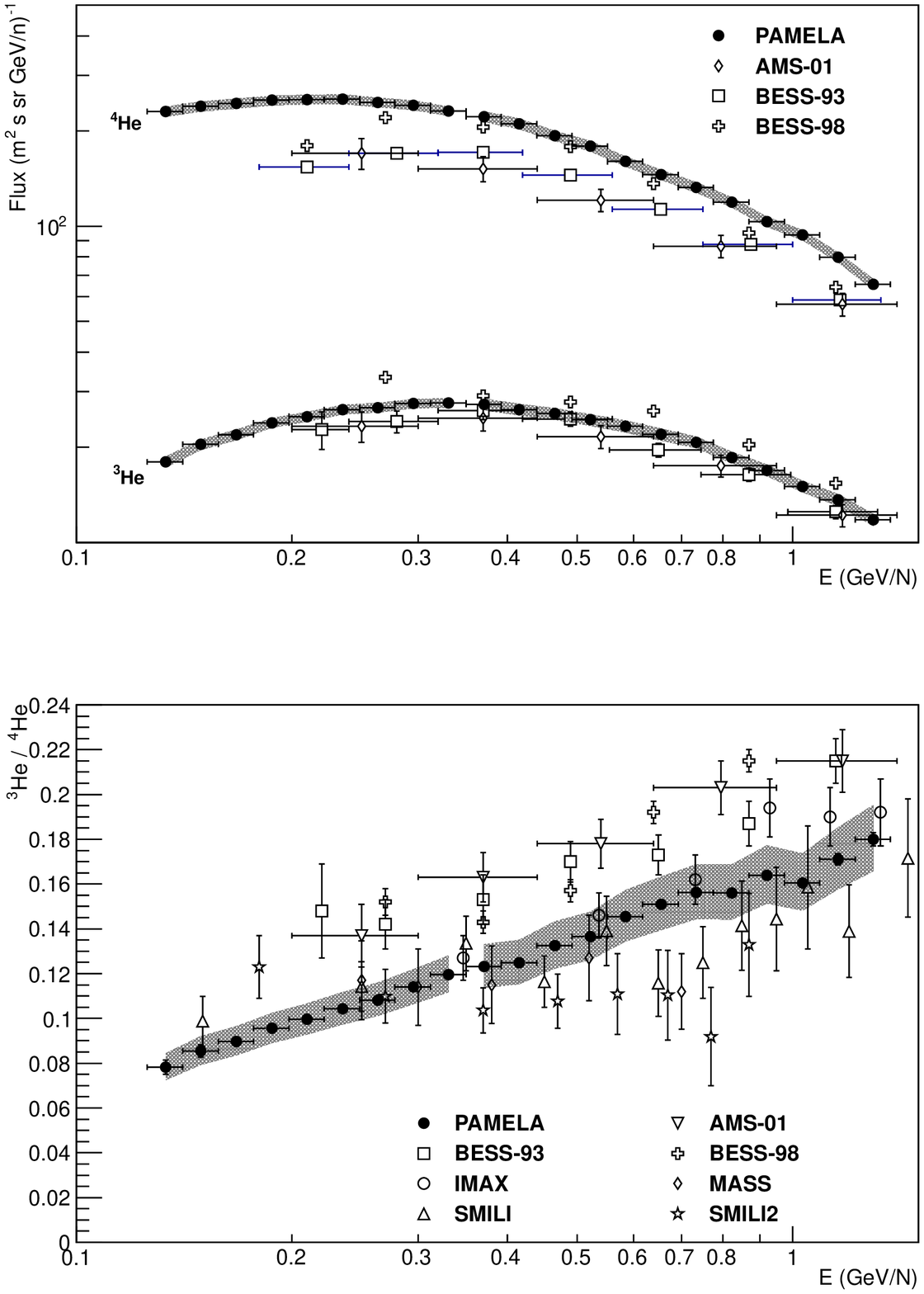}
    \caption{\hef\ and \het\ absolute fluxes (top) and their ratio (bottom).
For energies less than 350 MeV/n the ToF results (Table~\ref{tab:helium_tof}) were used, for higher energies the calorimeter results (Table~\ref{tab:helium_calo}). 
The previous experiments are: AMS \citep{2011ApJ...736..105A_red}, BESS-93 \citep{2002ApJ...564..244W}, 
BESS-98 \citep{2001ICRC...1805}, IMAX \citep{1998ApJ...496..490R_red}, SMILI-2\citep{1995ICRC....2..630W}, MASS \citep{1991ApJ...380..230W}, SMILI-1 \citep{1993ApJ...413..268B_red}. Error bars show statistical uncertainty while shaded areas show systematic uncertainty.}
    \label{im:fig_data_he2}
\end{figure*}

To compare our isotope fluxes with other measurements, we decided to use at low energies only the ToF results (up to 361 MeV/n for hydrogen and up to 350 MeV/n for helium) and above these values only use the calorimeter results. 
In Figure~\ref{im:fig_data_h2} and \ref{im:fig_data_he2} we show these hydrogen and helium isotope fluxes (top) and the ratios of the fluxes (bottom), compared to previous measurements~\citep{2002PhR...366..331A,2011ApJ...736..105A_red,2001ICRC...1617,2002ApJ...564..244W,
2005ASR...35..151,2001ICRC...1805,1998ApJ...496..490R_red,2000AIP...528..425,1995ICRC....2..630W,1991ApJ...380..230W,1993ApJ...413..268B_red}. 
 
\begin{figure*}[t]
    \centering
    \epsscale{0.75}
    \plotone{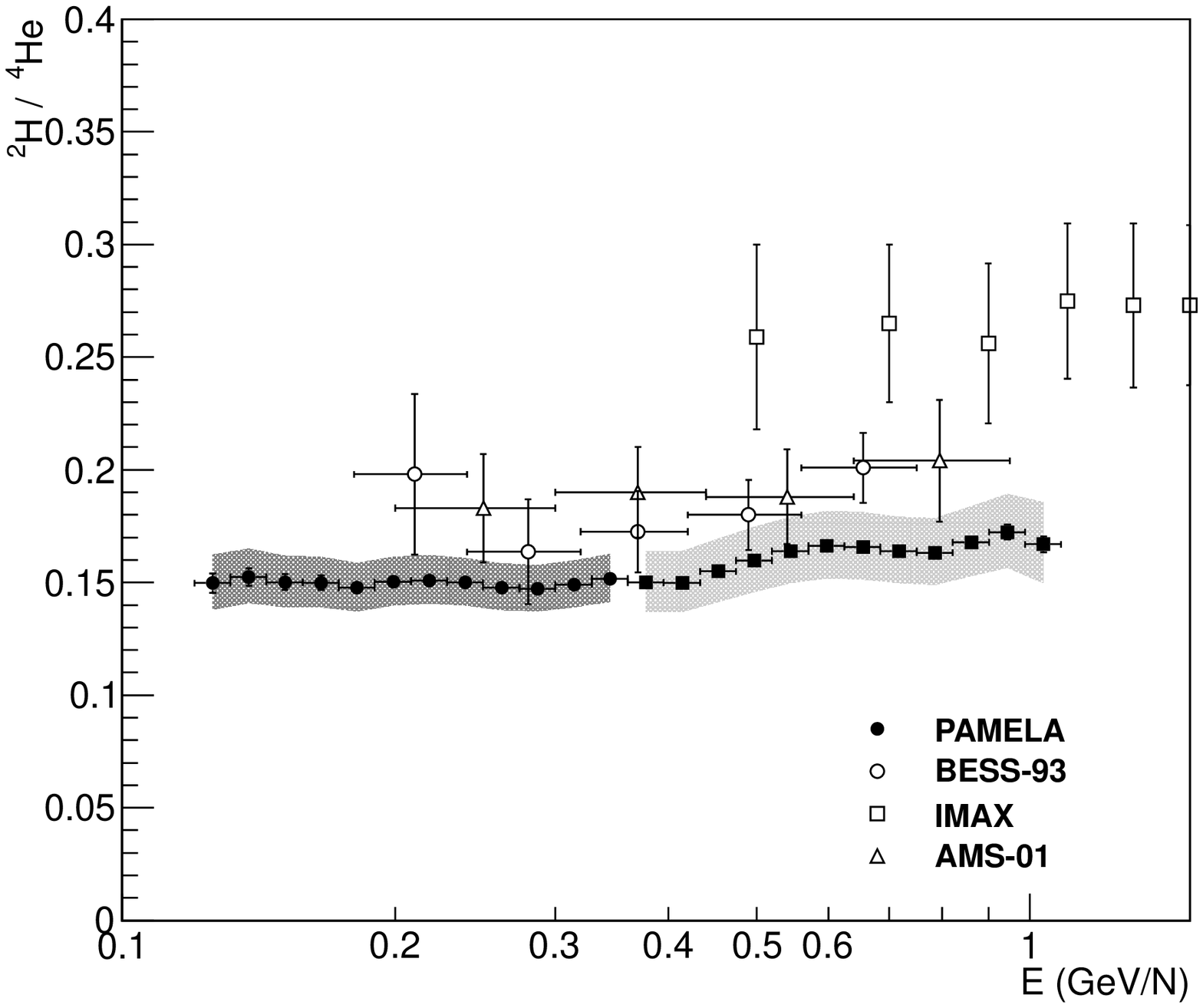}
    \caption{\deu/\hef\ ratio compared to previous experiments:  AMS-01 \citep{2011ApJ...736..105A_red}, BESS \citep{2002ApJ...564..244W}, 
IMAX \citep{2000AIP...528..425}. 
      Error bars show statistical uncertainty while shaded areas show systematic uncertainty. }
    \label{im:ratios}
\end{figure*}

Fig. \ref{im:ratios} shows the \deu/\hef\ ratio as a function of kinetic energy per nucleon.

It is visible that the former results show a large spread and it is obvious that the \pam\ results are more precise in terms of statistics. In this context it is important to know that all the former measurements shown in Figures~\ref{im:fig_data_h2}, \ref{im:fig_data_he2}, and \ref{im:ratios}, except AMS-01, are from balloon-borne experiments and thus  effected by the non-negligible background of atmospheric secondary particle production.

The scientific interest in these isotopes of \prot, \deu, \het\ and \hef\ are determined by the question about their origin. It is believed that the protons and the \hef\ particles are predominantly of primary origin thus arise directly from their sources while \deu\ and \het\ are of secondary origin thus are produced by interactions of these primaries with the interstellar gas. The interpretation of these results then allows to study more in detail the conditions of their propagation in the interstellar space. Beside these light isotopes presented here there are more particles of secondary origin which are used in these studies such as sub-iron particles or lithium, beryllium and boron. The effort aims to develop a diffusion model which will describe the propagation of charged particles and their lifetime in our galaxy. This will help also to better understand the energy density of different components within the interstellar space, such as magnetic fields, electromagnetic radiation, gas pressure and cosmic rays.  These model calculations have to deal with a number of parameters which have their origin in astrophysics, in nuclear physic and in high energy particle physics. The advantage of the light isotopes \deu\ and \het\ in this context compared to the more heavy secondary particles lies  in the fact that they do not have so many progenitor compared to the sub-iron particles or to lithium, beryllium and boron, it is predominantly \hef.
A comprehensive and detailed study and discussion and interpretation of our results in this context is beyond the scope of this paper but we like to refer to a recent paper published by \cite{2012A&A...539A..88C}.

\begin{deluxetable*}{cccc}
\tablecaption{Hydrogen isotope fluxes and their ratio derived with the ToF, errors are statistical and systematics respectively. \label{tab:hydrogen_tof}}
\tabletypesize{\scriptsize}
\tablewidth{0pt}
\tablehead{
\multicolumn{1}{c}{Kinetic energy} & \multicolumn{1}{c}{\prot\ flux} & \multicolumn{1}{c}{\deu\ flux} & \multicolumn{1}{c}{\deu / \prot} \\
\multicolumn{1}{c}{at top of payload} & \multicolumn{3}{c}{}\\
\multicolumn{1}{c}{(GeV n$^{-1}$)} & \multicolumn{1}{c}{(GeV n$^{-1}$ m$^2$ s sr)$^{-1}$} &  \multicolumn{1}{c}{(GeV n$^{-1}$ m$^2$ s sr)$^{-1}$} &
}
\startdata
0.120 - 0.132 & $(1.003 \pm 0.015 \pm 0.043) \cdot 10^{3}$ & $(33.9 \pm 0.9 \pm 1.5)$ & $(3.38 \pm 0.11 \pm 0.30) \cdot 10^{-2}$ \\ 
0.132 - 0.144 & $(1.062 \pm 0.014 \pm 0.044) \cdot 10^{3}$ & $(35.5 \pm 0.9 \pm 1.5)$ & $(3.34 \pm 0.09 \pm 0.28) \cdot 10^{-2}$ \\ 
0.144 - 0.158 & $(1.128 \pm 0.013 \pm 0.045) \cdot 10^{3}$ & $(36.0 \pm 0.8 \pm 1.4)$ & $(3.19 \pm 0.08 \pm 0.25) \cdot 10^{-2}$ \\ 
0.158 - 0.173 & $(1.186 \pm 0.012 \pm 0.045) \cdot 10^{3}$ & $(36.6 \pm 0.7 \pm 1.4)$ & $(3.08 \pm 0.07 \pm 0.24) \cdot 10^{-2}$ \\ 
0.173 - 0.190 & $(1.239 \pm 0.012 \pm 0.046) \cdot 10^{3}$ & $(36.7 \pm 0.7 \pm 1.4)$ & $(2.96 \pm 0.06 \pm 0.22) \cdot 10^{-2}$ \\ 
0.190 - 0.208 & $(1.296 \pm 0.011 \pm 0.047) \cdot 10^{3}$ & $(37.7 \pm 0.7 \pm 1.4)$ & $(2.91 \pm 0.06 \pm 0.21) \cdot 10^{-2}$ \\ 
0.208 - 0.228 & $(1.346 \pm 0.010 \pm 0.048) \cdot 10^{3}$ & $(38.0 \pm 0.6 \pm 1.4)$ & $(2.82 \pm 0.05 \pm 0.21) \cdot 10^{-2}$ \\ 
0.228 - 0.250 & $(1.406 \pm 0.010 \pm 0.050) \cdot 10^{3}$ & $(37.7 \pm 0.6 \pm 1.4)$ & $(2.68 \pm 0.04 \pm 0.19) \cdot 10^{-2}$ \\ 
0.250 - 0.274 & $(1.456 \pm 0.009 \pm 0.051) \cdot 10^{3}$ & $(36.5 \pm 0.5 \pm 1.3)$ & $(2.51 \pm 0.04 \pm 0.18) \cdot 10^{-2}$ \\ 
0.274 - 0.300 & $(1.487 \pm 0.009 \pm 0.052) \cdot 10^{3}$ & $(35.7 \pm 0.5 \pm 1.3)$ & $(2.40 \pm 0.03 \pm 0.17) \cdot 10^{-2}$ \\ 
0.300 - 0.329 & $(1.513 \pm 0.008 \pm 0.052) \cdot 10^{3}$ & $(35.2 \pm 0.4 \pm 1.2)$ & $(2.32 \pm 0.03 \pm 0.16) \cdot 10^{-2}$ \\ 
0.329 - 0.361 & $(1.513 \pm 0.007 \pm 0.052) \cdot 10^{3}$ & $(34.6 \pm 0.4 \pm 1.2)$ & $(2.29 \pm 0.03 \pm 0.16) \cdot 10^{-2}$ \\ 
0.361 - 0.395 & $(1.520 \pm 0.007 \pm 0.052) \cdot 10^{3}$ & $(33.6 \pm 0.4 \pm 1.2)$ & $(2.21 \pm 0.03 \pm 0.15) \cdot 10^{-2}$ \\ 
0.395 - 0.433 & $(1.499 \pm 0.006 \pm 0.051) \cdot 10^{3}$ & $(33.4 \pm 0.4 \pm 1.2)$ & $(2.23 \pm 0.03 \pm 0.16) \cdot 10^{-2}$ \\ 
0.433 - 0.475 & $(1.487 \pm 0.006 \pm 0.050) \cdot 10^{3}$ & $(32.3 \pm 0.4 \pm 1.2)$ & $(2.17 \pm 0.03 \pm 0.15) \cdot 10^{-2}$ \\ 
0.475 - 0.520 & $(1.469 \pm 0.006 \pm 0.050) \cdot 10^{3}$ & $(31.0 \pm 0.4 \pm 1.2)$ & $(2.11 \pm 0.03 \pm 0.15) \cdot 10^{-2}$ \\ 
0.520 - 0.570 & $(1.397 \pm 0.005 \pm 0.047) \cdot 10^{3}$ & $(29.4 \pm 0.3 \pm 1.3)$ & $(2.10 \pm 0.03 \pm 0.16) \cdot 10^{-2}$ \\ 
\enddata
\end{deluxetable*}

\begin{deluxetable*}{cccc}
\tablecaption{Hydrogen isotope fluxes and their ratio derived with the Calorimeter, errors are statistical and systematics respectively. \label{tab:hydrogen_calo}}
\tabletypesize{\scriptsize}
\tablewidth{0pt}
\tablehead{
\multicolumn{1}{c}{Kinetic energy} & \multicolumn{1}{c}{\prot\ flux} & \multicolumn{1}{c}{\deu\ flux} & \multicolumn{1}{c}{\deu / \prot} \\
\multicolumn{1}{c}{at top of payload} & \multicolumn{3}{c}{}\\
\multicolumn{1}{c}{(GeV n$^{-1}$)} & \multicolumn{1}{c}{(GeV n$^{-1}$ m$^2$ s sr)$^{-1}$} &  \multicolumn{1}{c}{(GeV n$^{-1}$ m$^2$ s sr)$^{-1}$} &
}
\startdata
0.228 - 0.250 & $(1.421 \pm 0.006 \pm 0.055) \cdot 10^{3}$ & $(37.0 \pm 0.7 \pm 2.0)$ & $(2.60 \pm 0.05 \pm 0.24) \cdot 10^{-2}$ \\ 
0.250 - 0.274 & $(1.474 \pm 0.005 \pm 0.057) \cdot 10^{3}$ & $(35.8 \pm 0.7 \pm 1.9)$ & $(2.43 \pm 0.05 \pm 0.22) \cdot 10^{-2}$ \\ 
0.274 - 0.300 & $(1.504 \pm 0.005 \pm 0.058) \cdot 10^{3}$ & $(36.0 \pm 0.6 \pm 1.9)$ & $(2.40 \pm 0.04 \pm 0.22) \cdot 10^{-2}$ \\ 
0.300 - 0.329 & $(1.528 \pm 0.005 \pm 0.058) \cdot 10^{3}$ & $(36.2 \pm 0.6 \pm 1.9)$ & $(2.37 \pm 0.04 \pm 0.22) \cdot 10^{-2}$ \\ 
0.329 - 0.361 & $(1.528 \pm 0.004 \pm 0.058) \cdot 10^{3}$ & $(35.2 \pm 0.6 \pm 1.9)$ & $(2.30 \pm 0.04 \pm 0.21) \cdot 10^{-2}$ \\ 
0.361 - 0.395 & $(1.534 \pm 0.004 \pm 0.058) \cdot 10^{3}$ & $(33.1 \pm 0.5 \pm 1.7)$ & $(2.16 \pm 0.03 \pm 0.19) \cdot 10^{-2}$ \\ 
0.395 - 0.433 & $(1.513 \pm 0.004 \pm 0.057) \cdot 10^{3}$ & $(31.7 \pm 0.5 \pm 1.7)$ & $(2.09 \pm 0.03 \pm 0.19) \cdot 10^{-2}$ \\ 
0.433 - 0.475 & $(1.498 \pm 0.003 \pm 0.056) \cdot 10^{3}$ & $(30.6 \pm 0.5 \pm 1.6)$ & $(2.04 \pm 0.03 \pm 0.18) \cdot 10^{-2}$ \\ 
0.475 - 0.520 & $(1.483 \pm 0.003 \pm 0.056) \cdot 10^{3}$ & $(29.6 \pm 0.4 \pm 1.6)$ & $(2.00 \pm 0.03 \pm 0.18) \cdot 10^{-2}$ \\ 
0.520 - 0.570 & $(1.441 \pm 0.003 \pm 0.054) \cdot 10^{3}$ & $(28.2 \pm 0.4 \pm 1.5)$ & $(1.96 \pm 0.03 \pm 0.18) \cdot 10^{-2}$ \\ 
0.570 - 0.625 & $(1.386 \pm 0.003 \pm 0.052) \cdot 10^{3}$ & $(26.2 \pm 0.4 \pm 1.4)$ & $(1.89 \pm 0.03 \pm 0.17) \cdot 10^{-2}$ \\ 
0.625 - 0.685 & $(1.314 \pm 0.002 \pm 0.049) \cdot 10^{3}$ & $(24.1 \pm 0.3 \pm 1.3)$ & $(1.84 \pm 0.03 \pm 0.17) \cdot 10^{-2}$ \\ 
0.685 - 0.750 & $(1.248 \pm 0.002 \pm 0.047) \cdot 10^{3}$ & $(22.2 \pm 0.3 \pm 1.2)$ & $(1.78 \pm 0.03 \pm 0.16) \cdot 10^{-2}$ \\ 
0.750 - 0.822 & $(1.179 \pm 0.002 \pm 0.044) \cdot 10^{3}$ & $(20.3 \pm 0.3 \pm 1.1)$ & $(1.73 \pm 0.03 \pm 0.16) \cdot 10^{-2}$ \\ 
0.822 - 0.901 & $(1.097 \pm 0.002 \pm 0.041) \cdot 10^{3}$ & $(18.9 \pm 0.3 \pm 1.0)$ & $(1.73 \pm 0.03 \pm 0.16) \cdot 10^{-2}$ \\ 
0.901 - 0.988 & $(1.0144 \pm 0.0018 \pm 0.0379) \cdot 10^{3}$ & $(17.5 \pm 0.3 \pm 1.0)$ & $(1.72 \pm 0.03 \pm 0.16) \cdot 10^{-2}$ \\ 
0.988 - 1.082 & $(9.216 \pm 0.016 \pm 0.343) \cdot 10^{2}$ & $(15.6 \pm 0.3 \pm 1.1)$ & $(1.70 \pm 0.03 \pm 0.18) \cdot 10^{-2}$ \\ 
\enddata
\end{deluxetable*}

\begin{deluxetable*}{cccc}
\tablecaption{Helium isotope fluxes and their ratio derived with the ToF, errors are statistical and systematics respectively. \label{tab:helium_tof}}
\tabletypesize{\scriptsize}
\tablewidth{0pt}
\tablehead{
\multicolumn{1}{c}{Kinetic energy} & \multicolumn{1}{c}{\hef\ flux} & \multicolumn{1}{c}{\het\ flux} & \multicolumn{1}{c}{\het / \hef} \\
\multicolumn{1}{c}{at top of payload} & \multicolumn{3}{c}{}\\
\multicolumn{1}{c}{(GeV n$^{-1}$)} & \multicolumn{1}{c}{(GeV n$^{-1}$ m$^2$ s sr)$^{-1}$} &  \multicolumn{1}{c}{(GeV n$^{-1}$ m$^2$ s sr)$^{-1}$} &
}
\startdata
0.126 - 0.141 & $(2.302 \pm 0.019 \pm 0.084) \cdot 10^{2}$ & $(18.0 \pm 0.7 \pm 0.7)$ & $(7.8 \pm 0.3 \pm 0.6) \cdot 10^{-2}$ \\ 
0.141 - 0.158 & $(2.392 \pm 0.017 \pm 0.085) \cdot 10^{2}$ & $(20.4 \pm 0.7 \pm 0.8)$ & $(8.5 \pm 0.3 \pm 0.6) \cdot 10^{-2}$ \\ 
0.158 - 0.177 & $(2.443 \pm 0.016 \pm 0.086) \cdot 10^{2}$ & $(21.9 \pm 0.6 \pm 0.8)$ & $(9.0 \pm 0.2 \pm 0.6) \cdot 10^{-2}$ \\ 
0.177 - 0.198 & $(2.501 \pm 0.015 \pm 0.087) \cdot 10^{2}$ & $(23.9 \pm 0.5 \pm 0.8)$ & $(9.6 \pm 0.2 \pm 0.7) \cdot 10^{-2}$ \\ 
0.198 - 0.222 & $(2.514 \pm 0.013 \pm 0.086) \cdot 10^{2}$ & $(25.0 \pm 0.5 \pm 0.9)$ & $(10.0 \pm 0.2 \pm 0.7) \cdot 10^{-2}$ \\ 
0.222 - 0.249 & $(2.522 \pm 0.012 \pm 0.086) \cdot 10^{2}$ & $(26.3 \pm 0.5 \pm 0.9)$ & $(1.043 \pm 0.019 \pm 0.072) \cdot 10^{-1}$ \\ 
0.249 - 0.279 & $(2.464 \pm 0.011 \pm 0.084) \cdot 10^{2}$ & $(26.7 \pm 0.4 \pm 0.9)$ & $(1.082 \pm 0.018 \pm 0.074) \cdot 10^{-1}$ \\ 
0.279 - 0.312 & $(2.413 \pm 0.010 \pm 0.082) \cdot 10^{2}$ & $(27.5 \pm 0.4 \pm 0.9)$ & $(1.140 \pm 0.017 \pm 0.078) \cdot 10^{-1}$ \\ 
0.312 - 0.350 & $(2.310 \pm 0.009 \pm 0.078) \cdot 10^{2}$ & $(27.6 \pm 0.4 \pm 0.9)$ & $(1.196 \pm 0.017 \pm 0.081) \cdot 10^{-1}$ \\ 
0.350 - 0.392 & $(2.226 \pm 0.009 \pm 0.075) \cdot 10^{2}$ & $(27.6 \pm 0.3 \pm 0.9)$ & $(1.239 \pm 0.016 \pm 0.084) \cdot 10^{-1}$ \\ 
0.392 - 0.439 & $(2.105 \pm 0.008 \pm 0.071) \cdot 10^{2}$ & $(27.1 \pm 0.3 \pm 0.9)$ & $(1.289 \pm 0.016 \pm 0.087) \cdot 10^{-1}$ \\ 
0.439 - 0.492 & $(1.922 \pm 0.007 \pm 0.065) \cdot 10^{2}$ & $(26.3 \pm 0.3 \pm 0.9)$ & $(1.367 \pm 0.016 \pm 0.093) \cdot 10^{-1}$ \\ 
0.492 - 0.551 & $(1.802 \pm 0.006 \pm 0.061) \cdot 10^{2}$ & $(25.2 \pm 0.3 \pm 0.9)$ & $(1.400 \pm 0.015 \pm 0.096) \cdot 10^{-1}$ \\ 
0.551 - 0.618 & $(1.669 \pm 0.006 \pm 0.056) \cdot 10^{2}$ & $(24.8 \pm 0.2 \pm 0.9)$ & $(1.488 \pm 0.016 \pm 0.103) \cdot 10^{-1}$ \\ 
0.618 - 0.692 & $(1.536 \pm 0.005 \pm 0.052) \cdot 10^{2}$ & $(23.7 \pm 0.2 \pm 0.9)$ & $(1.543 \pm 0.016 \pm 0.110) \cdot 10^{-1}$ \\ 
0.692 - 0.776 & $(1.383 \pm 0.005 \pm 0.048) \cdot 10^{2}$ & $(22.2 \pm 0.2 \pm 0.9)$ & $(1.602 \pm 0.016 \pm 0.119) \cdot 10^{-1}$ \\ 
0.776 - 0.870 & $(1.244 \pm 0.004 \pm 0.045) \cdot 10^{2}$ & $(19.81 \pm 0.19 \pm 0.87)$ & $(1.593 \pm 0.016 \pm 0.127) \cdot 10^{-1}$ \\ 
\enddata
\end{deluxetable*}

\begin{deluxetable*}{cccc}
\tablecaption{Helium isotope fluxes and their ratio derived with the Calorimeter, errors are statistical and systematics respectively. \label{tab:helium_calo}}
\tabletypesize{\scriptsize}
\tablewidth{0pt}
\tablehead{
\multicolumn{1}{c}{Kinetic energy} & \multicolumn{1}{c}{\hef\ flux} & \multicolumn{1}{c}{\het\ flux} & \multicolumn{1}{c}{\het / \hef} \\
\multicolumn{1}{c}{at top of payload} & \multicolumn{3}{c}{}\\
\multicolumn{1}{c}{(GeV n$^{-1}$)} & \multicolumn{1}{c}{(GeV n$^{-1}$ m$^2$ s sr)$^{-1}$} &  \multicolumn{1}{c}{(GeV n$^{-1}$ m$^2$ s sr)$^{-1}$} &
}
\startdata
0.249 - 0.279 & $(2.501 \pm 0.014 \pm 0.094) \cdot 10^{2}$ & $(28.9 \pm 0.6 \pm 1.2)$ & $(1.15 \pm 0.02 \pm 0.09) \cdot 10^{-1}$ \\ 
0.279 - 0.312 & $(2.411 \pm 0.013 \pm 0.091) \cdot 10^{2}$ & $(29.8 \pm 0.6 \pm 1.2)$ & $(1.24 \pm 0.02 \pm 0.10) \cdot 10^{-1}$ \\ 
0.312 - 0.350 & $(2.292 \pm 0.012 \pm 0.086) \cdot 10^{2}$ & $(28.8 \pm 0.5 \pm 1.2)$ & $(1.26 \pm 0.02 \pm 0.10) \cdot 10^{-1}$ \\ 
0.350 - 0.392 & $(2.222 \pm 0.011 \pm 0.083) \cdot 10^{2}$ & $(27.4 \pm 0.5 \pm 1.1)$ & $(1.23 \pm 0.02 \pm 0.10) \cdot 10^{-1}$ \\ 
0.392 - 0.439 & $(2.110 \pm 0.010 \pm 0.079) \cdot 10^{2}$ & $(26.3 \pm 0.4 \pm 1.1)$ & $(1.25 \pm 0.02 \pm 0.10) \cdot 10^{-1}$ \\ 
0.439 - 0.492 & $(1.931 \pm 0.009 \pm 0.072) \cdot 10^{2}$ & $(25.6 \pm 0.4 \pm 1.1)$ & $(1.32 \pm 0.02 \pm 0.10) \cdot 10^{-1}$ \\ 
0.492 - 0.551 & $(1.791 \pm 0.008 \pm 0.067) \cdot 10^{2}$ & $(24.4 \pm 0.3 \pm 1.0)$ & $(1.36 \pm 0.02 \pm 0.11) \cdot 10^{-1}$ \\ 
0.551 - 0.618 & $(1.601 \pm 0.007 \pm 0.059) \cdot 10^{2}$ & $(23.3 \pm 0.3 \pm 1.0)$ & $(1.45 \pm 0.02 \pm 0.11) \cdot 10^{-1}$ \\ 
0.618 - 0.692 & $(1.455 \pm 0.007 \pm 0.054) \cdot 10^{2}$ & $(22.0 \pm 0.3 \pm 0.9)$ & $(1.51 \pm 0.02 \pm 0.12) \cdot 10^{-1}$ \\ 
0.692 - 0.776 & $(1.327 \pm 0.006 \pm 0.049) \cdot 10^{2}$ & $(20.7 \pm 0.3 \pm 0.9)$ & $(1.56 \pm 0.02 \pm 0.12) \cdot 10^{-1}$ \\ 
0.776 - 0.870 & $(1.190 \pm 0.006 \pm 0.044) \cdot 10^{2}$ & $(18.6 \pm 0.2 \pm 0.8)$ & $(1.56 \pm 0.02 \pm 0.12) \cdot 10^{-1}$ \\ 
0.870 - 0.974 & $(1.034 \pm 0.005 \pm 0.038) \cdot 10^{2}$ & $(16.9 \pm 0.2 \pm 0.7)$ & $(1.64 \pm 0.02 \pm 0.13) \cdot 10^{-1}$ \\ 
0.974 - 1.092 & $(93.8 \pm 0.5 \pm 3.5)$ & $(15.1 \pm 0.2 \pm 0.6)$ & $(1.60 \pm 0.02 \pm 0.13) \cdot 10^{-1}$ \\ 
1.092 - 1.223 & $(79.7 \pm 0.4 \pm 3.0)$ & $(13.63 \pm 0.20 \pm 0.59)$ & $(1.71 \pm 0.03 \pm 0.14) \cdot 10^{-1}$ \\ 
1.223 - 1.371 & $(65.5 \pm 0.4 \pm 2.4)$ & $(11.80 \pm 0.18 \pm 0.52)$ & $(1.80 \pm 0.03 \pm 0.15) \cdot 10^{-1}$ \\ 
\enddata
\end{deluxetable*}

\section{Acknowledgments}
We acknowledge support from the Russian Space Agency (Roscosmos), the Russian Foundation for Basic Research (grant 13-02-00298), the Russian Scientific Foundation (grant 14-12-00373), the Italian Space Agency (ASI), Deutsches Zentrum fur Luft- und Raumfahrt (DLR), the Swedish National Space Board, and the Swedish Research Council.

\end{document}